\newcommand{\arcsec}{\mbox{$^{\prime\prime$}}}

\documentclass[]{aa}
\usepackage{graphicx,latexsym,amssymb}
\usepackage{lscape}
\usepackage[dvips]{rotating}

\begin{document}

   \title{Accurate photometry of extended spherically symmetric sources}

   \author{P. Anders
          \inst{1}
          \and
          M. Gieles\inst{2}
           \and
          R. de Grijs\inst{3}
         }

   \offprints{P. Anders}

   \institute{Institut f\"ur Astrophysik, Friedrich-Hund-Platz 1, 37077 G\"ottingen, Germany\\
              \email{panders@astro.physik.uni-goettingen.de}
         \and
     Astronomical Institute, Utrecht University, Princetonplein 5, 3584 CC Utrecht, The Netherlands\\
     \email{M.Gieles@astro.uu.nl}
         \and
     Department of Physics \& Astronomy, The University of Sheffield, Hicks Building, Hounsfield Road, Sheffield, S3 7R, UK\\
     \email{R.DeGrijs@sheffield.ac.uk}
              }

\date{Received xxx, Accepted xxx}

\abstract{We present a new method to derive reliable photometry of extended
spherically symmetric sources from {\it HST} images (WFPC2, ACS/WFC and NICMOS/NIC2
cameras), extending existing studies of point sources and marginally resolved
sources. We develop a new approach to accurately determine intrinsic
sizes of extended spherically symmetric sources, such as star clusters in galaxies beyond
the Local Group (at distances $\lesssim 20$ Mpc), and provide a
detailed cookbook to perform aperture photometry on such sources, by
determining size-dependent aperture corrections (ACs) and taking sky
oversubtraction as a function of source size into account.\\ 
In an extensive Appendix, we provide the parameters of polynomial
relations between the FWHM of various input profiles and those
obtained by fitting a Gaussian profile (which we have used for reasons
of computational robustness, although the exact model profile used is
irrelevant), and between the intrinsic and measured FWHM of the
cluster and the derived AC. Both relations are given for a number of
physically relevant cluster light profiles, intrinsic and
observational parameters. AC relations are provided for a wide range
of apertures. Depending on the size of the source and the annuli used
for the photometry, the absolute magnitude of such extended objects can be
underestimated by up to 3 mag, corresponding to an error in mass of a
factor of 15.\\
We carefully compare our results to those from the more widely used
DeltaMag method, and find an improvement of a factor of 3--40 in both
the size determination and the AC.
}


\maketitle

\section{Introduction}

We present a new method to determine photometric properties of 
extended spherically symmetric sources in {\sl Hubble Space Telescope (HST)} data obtained
with the Wide Field and Planetary Camera 2 (WFPC2), the Advanced
Camera for Surveys / Wide Field Camera (ACS/WFC), and camera 2 of
the Near Infrared Camera and Multi Object Spectrometer (NICMOS).

When studying extragalactic star clusters (SCs) at high spatial
resolution, such as with the {\it HST}, the accuracy of ``classical''
photometric methods becomes insufficient. Ideally, fitting the
point-spread functions (PSFs) is desirable for sources in crowded
fields and with variable background fluxes. However, this is difficult
since SCs at distances of $\lesssim$ 20 Mpc appear extended on the
{\sl HST} images and, as a consequence, PSF fitting techniques will
underestimate their true fluxes.

With the best spatial resolution possible to achieve today ($\sim0.05$
arcsec with the {\sl HST}, namely using WFPC2/PC, ACS/WFC and ACS/HRC) many 
nearby clusters are clearly resolved. We
define a ``clearly resolved'' cluster conservatively as having 1.2
$\times$ the PSF size, and hence an {\sl observed} cluster FWHM roughly of
the order of 2.3 pixels (see Table \ref{tab:psfsizes}). As will be
shown below, these 2.3 pixel correspond to an {\sl intrinsic} cluster FWHM 
on the order of 0.5 pixel.

In addition, the high spatial resolution of the WFPC2 and ACS
cameras undersample the PSF. For {\it marginally} extended sources,
a satisfactory solution to this undersampling problem has recently
been included in the {\sc HSTphot} PSF fitting software package
custom-written to handle {\sl HST} photometry (Dolphin 2000).

Measuring the light in a fixed annulus around the central source
coordinates, as commonly done in aperture photometry, can in
principle correct for both the undersampled PSF and source size.
However, when studying a population of sources with variable sizes,
as for extragalactic SC systems in general, using a fixed aperture
will underestimate the flux of the larger sources with respect to
that of the point-like sources.

Many extragalactic SC studies have tried to estimate the size of the
sources based on the magnitude difference in different apertures (we
will refer to this as the ``DeltaMag method''), and compare these to
either model clusters (usually assuming Gaussian light profiles;
e.g. Whitmore et al. 1993; Whitmore \& Schweizer 1995; Zepf et al.
1999) and/or observed star profiles (e.g. Zepf et al. 1999).
Sometimes, multiple apertures and cumulative light distributions are
used, thus enhancing reliability (e.g. Puzia et al. 1999). However, as
shown in de Grijs et al. (2001), the presence of a variable,
structured background strongly compromises the results from the
DeltaMag method.

In the same studies, estimates of aperture corrections (ACs) needed
to account for the finite size of the objects are given, again on
the basis of either model clusters (e.g. Whitmore \& Schweizer 1995)
or isolated clusters in the science images of interest (e.g. Miller
et al. 1997; Carlson et al. 1998), mostly determined for a subset of
clusters and applied to the whole sample -- independent of object
size. Some authors do attempt to use size-dependent ACs (e.g. Zepf
et al. 1999), although generally not well defined, and mostly based
on the rough size estimates resulting from the magnitude difference
method. This method is vulnerable to centering problems (the use of
0.5 pixel radius apertures is seen regularly), and the sizes (and
derived size-dependent ACs, as a consequence) are only rough
estimates.

Other studies are based on more subjective methods, such as those
that determine the source and sky annuli for each cluster
individually, to encircle the dominant cluster light contribution
and to avoid background contamination (see e.g. de Grijs et al.
2001; Anders et al. 2004). While this method avoids ACs (since it is 
already supposed to measure the dominant
light contribution), it is hampered by subjectivity, and does not
provide reliable size estimates.

There exist, as yet, no large-scale theoretical studies of the
reliability, reproducibility and comparability of the results for any
of these methods. All are subject to subjectivity in one aspect or
another (e.g. the choice of apertures for size estimates/photometry,
cluster light profile, selection of a few single clusters to derive
``average'' ACs).

To date, only two sophisticated systematic studies  have been done
to determine accurate SC sizes: 
\begin{itemize}

\item Carlson \& Holtzman (2001), but limited to marginally
resolved, high S/N sources, without studying the accompanying ACs

\item Dolphin \& Kennicutt (2002) related to the above-mentioned
program package HSTphot and its application to (again) marginally
resolved sources in NGC 3627. This study is based on a PSF-fitting
strategy for extended sources, while our work is based on
aperture photometry.

\end{itemize}

The present study complements, expands upon and enhances those of
Carlson \& Holtzman (2001) and Dolphin \& Kennicutt (2002). This
study also fully complements structural studies of resolved
clusters, e.g. in the Large and Small Magellanic Clouds (LMC, SMC)
and nearby dwarf galaxies (see e.g. Mackey \& Gilmore 2003a,b).
However, such studies are only possible for the very nearest
galaxies and their clusters. Where ACs are concerned,  this study
extends the widely-used work of Holtzman et al. (1995) for point
sources to the studies of extended spherically symmetric sources.

In this paper, we present a new method to perform more accurate
aperture photometry of extended spherically symmetric sources using a simple extension to
the basic principle of aperture photometry. After measuring the flux
of each source using a fixed aperture size, a variable AC
 (based on the actual size of the object) is applied. This
method greatly enhances reproducibility and comparability of the
results obtained. With the large range of parameter space explored
and numerous related effects taken into account, we also present for
the first time a method to estimate uncertainties in the sizes and
ACs for a given observation.

In Section \ref{sec:size} we propose a general definition of
``size'', as a function of a large number of intrinsic and
observational parameters. In Section \ref{sec:ac} the relation
between source FWHM and the appropriate AC is
determined as a function of aperture size. In Section
\ref{sec:cookbook} we provide a detailed ``cookbook'', ready for
immediate application to extragalactic SC systems. The reader who is
only interested in applying our ACs could skip
directly to Section \ref{sec:cookbook}.  In Section
\ref{sec:comparison} we provide an example error analysis for our new
method, including a comparison of the method presented here to the
DeltaMag method.

\section{Determining accurate source sizes}
\label{sec:size}

Conventionally, the stellar density distributions of old globular
clusters (GCs) are well described by King profiles (King 1962)
with a range of concentration parameters. Intermediate-age and
young star clusters (YSCs), e.g., in the LMC, are better
described by Elson, Fall \& Freeman (1987; EFF) profiles. Such
clusters, similar to YSCs in, e.g., the Antennae galaxies or NGC
7252, do not (yet) show evidence of tidal truncations, in
contrast to King profiles. We set out to analyse SC systems
containing SCs spanning a large range of ages, masses and sizes,
and compare radii and compactnesses of SCs in different galaxies.
We therefore need a reliable method to estimate, to high
accuracy, the radii of a large variety of SCs.

Thus, we first have to find a general definition of ``size''. To
this end, we created artificial SCs based on a variety of profiles
using the {\sc BAOlab} package of Larsen (1999) (for recent
applications see Larsen 2004a; Boeker et al. 2004). {\sc BAOlab}
creates artificial clusters of a given magnitude by randomly
drawing the position of each recorded photon from the input light
profile. It thus simulates the stochastic nature of real
observations very effectively, indeed more effectively than any
other program available. These profiles were convolved with
pre-calculated PSFs, generated with the {\sc Tiny Tim} package
(Krist \& Hook 2004), and (for WFPC2 and ACS/WFC) the appropriate
diffusion kernels supplied by {\sc Tiny Tim}. Although some caveats
still exist, {\sc Tiny Tim} is the best suited package to obtain
realistic {\sl HST} PSFs to date. First, it extensively covers the 
parameter space of interest (cameras, filters, chips, position on 
the chips, object spectra, focus, PSF sizes
etc.), well beyond anything that can be realisticly done with observed
PSFs. And secondly, and even more crucial, the subsampling of 
{\sc Tiny Tim} PSFs allows one to study the real distribution of counts
onto adjacent pixels, depending on the exact PSF peak position on
subpixel basis. This subsampling is fully implemented and used in
the {\sc BAOlab} package.

In order to measure the size of these objects
we fit Gaussian profiles to them. Many extragalactic SC systems
observed to date display a wide range of cluster sizes, so that we
need to have a consistent and robust size determination to compare
SC sizes and compactnesses. Therefore, we decided to apply a
blanket fitting approach of Gaussian profiles to the SC light
distributions. We realise that this is a simplification, but
fitting more complicated profiles (such as King or EFF profiles)
requires high signal-to-noise (S/N) ratios and the knowledge of
whether or not the clusters are tidally truncated. In practice,
this will be difficult for a large number of sources in realistic
SC samples. We point out that the actual, underlying cluster
profile is only of minor importance for the {\it relative} size
determinations of SCs in a given SC system; the key prerogative is
that one applies a consistent approach to one's size
determinations. Since we also base our ACs on such
Gaussian fits our approach is fully internally consistent, and we
have, in effect, taken the detailed profile shape out of the
equation. In Sections \ref{sec:size_standard} to
\ref{sec:size_filter} we will perform a detailed analysis for the
WFPC2 camera. In Section \ref{sec:size_acs} and
\ref{sec:size_nic2} we will expand this to the ACS/WFC and
NICMOS/NIC2 cameras.

Finally, we note that fitting more realistic light profiles
results in less stable fit results, since either King or EFF
profiles have one additional free parameter (concentration and
power-law slope, respectively). They are also more sensitive to
features at the periphery of the cluster, i.e., in the low-S/N
regions.

Throughout this paper, we will use the FWHM of the input profile and
the measured FWHM of the fitted Gaussian profile as measures for the
size. In Table \ref{tab:size_con} we present the (constant) conversion
factors from FWHM to the more widely used half-light radius,
$R_{1/2}$, for the different models (e.g., Larsen 2004b).

\subsection{The parameters of the ``standard'' cluster} 
\label{sec:size_standard}

In the following subsections we will investigate the behaviour of the
measured cluster sizes (using the FWHM of a Gaussian profile, as
justified in the previous section) as a function of the input FWHM,
assuming various parameters for the artificial clusters and a range of
observational conditions. In the Appendices we provide conversion
relations between input and measured FWHM (and vice versa), by fitting
fifth-order polynomials to these conversion relations, of the form
\begin{equation}
{\rm size}(x) = a + b*x + c*x^2 + d*x^3 + e*x^4 + f*x^5
\end{equation}
and
\begin{equation}
{\rm size}'(y) = a' + b'*y + c'*y^2 + d'*y^3 + e'*y^4 + f'*y^5
\end{equation}
where $x$ and size$'(y)$ are the intrinsic FWHM in pixels, and size$(x)$
and $y$ the measured FWHM. We decided to use fifth-order polynomials
after a visual inspection of the data and the fit results, as a
compromise between fitting details in the shape, wiggles and instabilities in the fits,
and usability. We note that this choice is purely based on
mathematical convenience, and not on any physical properties of the
SCs.

We first define our ``standard'' cluster:

\begin{itemize}

\item {\bf ``Observations''}: cluster of mag(WFPC2/F555W $\approx V) 
= 10$ mag, observed in a 1s exposure;

\item {\bf {\sc Tiny Tim} PSF properties}: {\sl HST} WFPC2/WF3
chip;  central position on the chip, i.e., $(x,y) = (400,400)$;
F555W filter; using standard WFPC2 diffusion kernel;

\item {\bf {\sc BAOlab} parameters}: no noise, profile fitting
radius = 5 pixels; used for cluster generation and cluster fitting;

\item {\bf Input cluster light profiles}: Gaussian; King (1962) 
models with concentrations $c = 5, 30, 100$ \footnote{The
concentration parameter, $c$, is the ratio of tidal to core
radius, $c \equiv r_t/r_c$; note that the concentration is more
often given as $\log(c)$}; Elson et al. (1987) models with
power-law index $\gamma = 1.5, 2.5$ (cf. Section
\ref{sec:size_models});

\item {\bf Fit model}: Two-dimensional (2D) Gaussian profile,
without  taking into account any {\sc Tiny Tim} {\sl HST} PSF or
{\sl HST} diffusion kernel (effectively using a delta
function-type PSF). \end{itemize}

Although in the following sections we will plot the conversion
relations for a larger range of input FWHMs (in order to illustrate
that we understand their behaviour across the entire range of
realistic sizes), we strongly advise to use these relations only in
the range of $0.5 \le $ input FWHM (pixels) $\le 10$. For smaller
input FWHM, the data are not well approximated by the fitted
polynomials. For larger input FWHMs, the S/N ratio per pixel
decreases, and as a consequence the noise increases, so that the fit
will not be sufficiently accurate. In other words, after converting
measured radii to ``intrinsic'' radii, it is advisable to treat
clusters with ``intrinsic'' radii outside the 0.5 -- 10 pixel range
with caution.

\subsection{Size determination as a function of input model}
\label{sec:size_models}

In this Section we use the ``standard'' clusters defined in Section
\ref{sec:size_standard}. The models used are a Gaussian model,
King (1962) models with $c =$ 5, 30 and 100 (King 5, King 30 and
King 100, respectively), and EFF profiles (Elson et al. 1987) with
power-law indices $\gamma =$ 1.5 and 2.5 (EFF 15 and EFF 25,
respectively). We point out that the power-law index $\gamma$ used in
{\sc BAOlab} differs by a factor of 2 from the definition used by
Elson et al. (1987), with $\gamma_{\rm EFF}=2\times \gamma_{\rm
BAOlab}$.

The light profiles are represented by the following equations, where
$r$ is the (dimensionless) radius (in units of FWHM), and $w$ is a 
(dimensionless) normalisation constant:

\begin{itemize}
\item Gaussian:
\begin{equation}
f(r)=\exp \left (- \left(r \cdot w\right)^2\right),
\end{equation}
with $w=2 \sqrt{\ln(2)} \approx 1.66$

\item King models:
\begin{equation}
f(r,c)=\left(\frac{1}{\sqrt{1+\left(r \cdot
w\right)^2}}-\frac{1}{\sqrt{1+c^2}}\right)^2,
\end{equation}
with $w=2 \sqrt{\left( \sqrt{0.5} + \frac{1-\sqrt{0.5}}{\sqrt{1+c^2}}
\right)^{-2} -1}$, so that $w\approx 1.69, 1.95$ and 1.98 for King 5,
King 30 and King 100, respectively.

\item EFF models:
\begin{equation}
f(r,\gamma)=\left(1+\left(r \cdot w\right)^2\right)^{-\gamma},
\end{equation}
with $w=2 \sqrt{2^{1/\gamma}-1}$, i.e., $w\approx1.53$ and 1.13 for
EFF 15 and EFF 25, respectively.
\end{itemize}     

\begin{table}

\caption{Conversion factors to calculate a model's $R_{\rm eff}=R_{1/2}$ from
its FWHM.}

\begin{center}
\begin{tabular}{@{}*{2}{|c}{|}@{}}
\hline
model & $R_{1/2}$/FWHM \\
\hline
GAUSS   & 0.5  \\
King 5   & 0.71 \\
King 30  & 1.48 \\
King 100 & 2.56 \\
EFF 15   & 1.13 \\
EFF 25   & 0.68 \\
\hline
\end{tabular}
\label{tab:size_con}
\end{center}
\end{table}

\subsubsection{Effect of the PSF on a Gaussian profile}
\label{subsec:size_gauss}

The first step is to assess what the effect of PSF ``blurring'' is on
Gaussian profiles. Standard clusters (see Section
\ref{sec:size_standard}) were created using Gaussian input profiles
with different FWHMs. Subsequently, 2D Gaussian profiles were fit to
the resulting images. A Gaussian fit results in either a Gaussian
width, $\sigma$, where FWHM $= 2\sqrt{2\ln(2)} \sigma$, or directly in
the FWHM, in most commonly used Gaussian fitting routines.

\begin{figure}
    \begin{center}
    \includegraphics[angle=180,width=0.75\columnwidth]{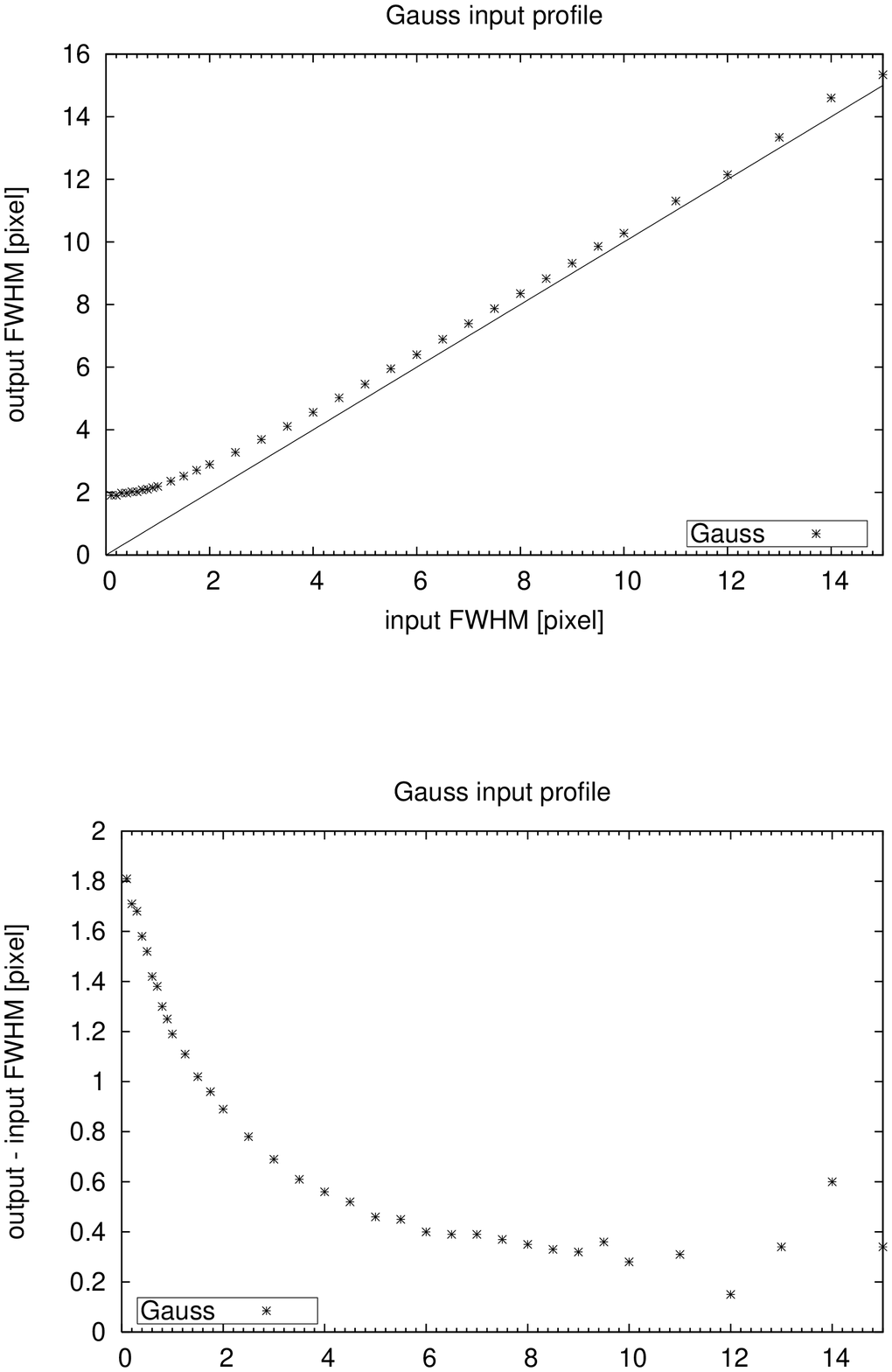}
    \end{center} 
    \vspace{0.6cm}
    \caption{Fitted Gaussian FWHMs for input Gaussian profile
    convolved with WFPC2/WF3 F555W PSF, located on the central pixel.
    The diagonal solid line in the top panel represents a one-to-one
    relation, the other solid lines are the fifth-order polynomial
    fits to the data. Top: Output FWHM. Bottom: Output $-$ input FWHM.}
    \label{fig:size_gauss} 
    \begin{center}
    \includegraphics[angle=270,width=0.95\columnwidth]{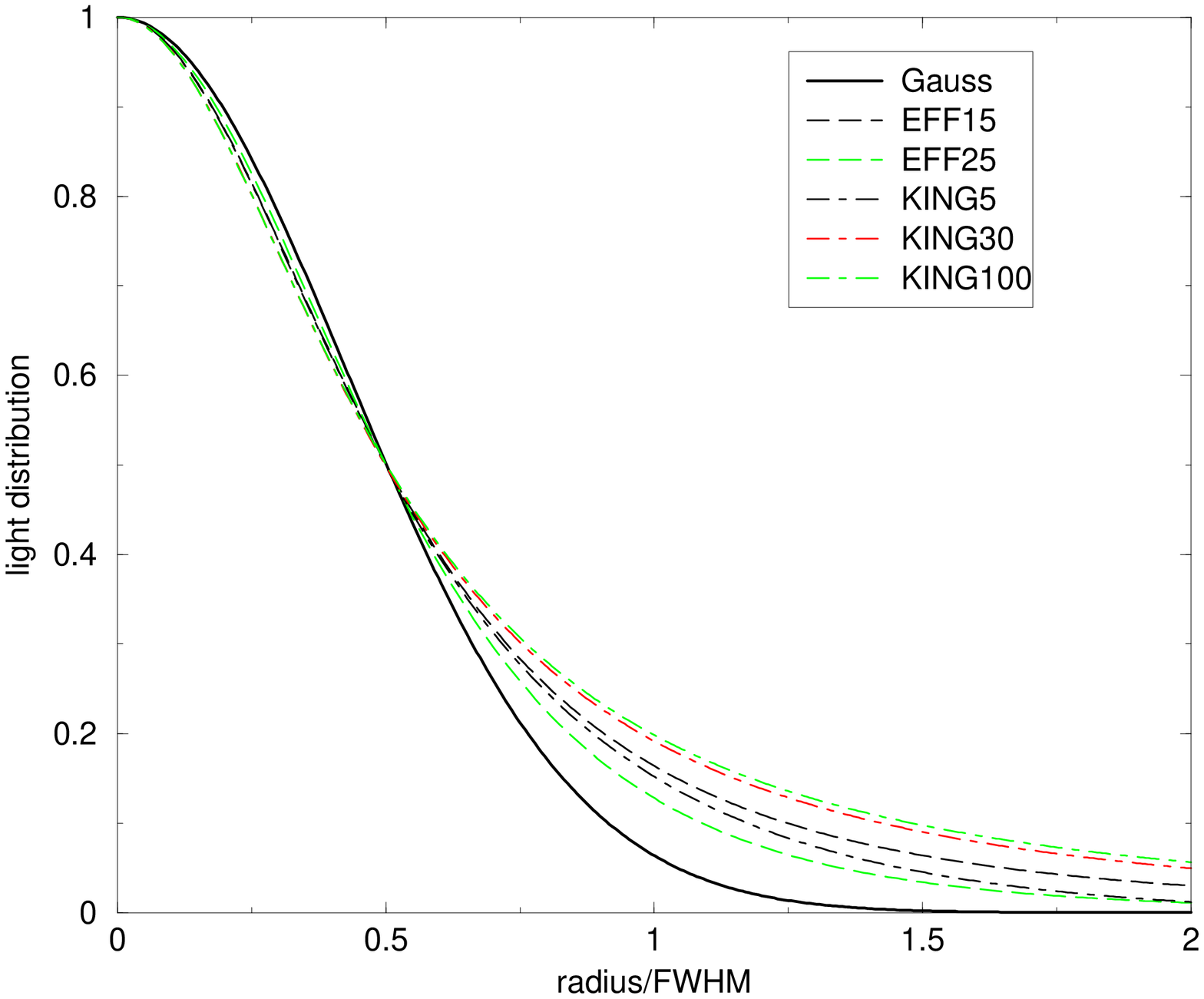}
    \end{center}
    \caption{Model light distributions.} 
    \label{fig:size_models2}
\end{figure}

Results for a range of input FWHM values from 0.1 to 15 WF3 pixels are
shown in Fig.~\ref{fig:size_gauss}. The offset caused by the
convolution of the input profile with the instrumental PSF decreases
with increasing input cluster size, since the PSF and diffusion kernel
broadening become less and less important. For clusters with input
FWHMs greater than $\sim 3$ pixels, the relation between input FWHM
and recovered FWHM of the Gaussian fit is approximately linear, and
the derived (``measured'') cluster sizes are of the order of 0.3--0.6
pixels (3--20 per cent) larger than the input (``intrinsic'') values.

For clusters with input FWHMs greater than $\sim 10$ pixels, the
scatter increases because of the low S/N ratio per pixel.

To conclude, we understand the general behaviour of this data set very
well. However, since the Gaussian cluster light profile is the least
realistic input profile, we will not consider it in the remainder of
this study. In this section we simply wanted to demonstrate that the
method works in a comprehensible way. In the following sections we
will use more realistic input light profiles.

\subsubsection{Non-Gaussian input models}
\label{sec:size_remarks}

For a given FWHM, the less concentrated King/EFF model profiles have less
light in the wings than more concentrated King/EFF profiles, while a comparable 
Gaussian profile has least flux in the wings. At radii smaller than the FWHM,
non-Gaussian models appear somewhat more compact than a Gaussian profile. This
is illustrated in Fig.~\ref{fig:size_models2}.

In the first part of this study we adopt a fixed fitting radius of 5
pixels. We made this conscious choice for the fitting radius because
for any realistic extragalactic SC observed with the {\sl HST} at a
decent S/N ratio, profile fits using Gaussian profiles are generally
feasible. Larger fitting radii may be unproportionally affected by
non-Gaussianity in the cluster profiles, low-S/N regions (i.e.,
fluctuations in the background noise), or neighbouring objects in
crowded regions; much smaller fitting radii may not always be
appropriate to employ Gaussian profile fits. To illustrate this, in
Section \ref{sec:size_fitrad}, we will show that changing the fitting
radius leads to systematic changes (and even numerical instabilities)
of the ACs, and explain why this is the case.

Combining the choice of our 5-pixel fitting radius and the general
behaviour of our input models, we expect to see the following trends:

\begin{itemize}
\item For clusters with input FWHM greater than 5 pixels, only the 
inner core will be fit. Due to the greater compactness of non-Gaussian
models with respect to Gaussian profiles, for a given FWHM, we expect
to systematically underestimate the sizes of large clusters.

\item For clusters with FWHM smaller than 5 pixels, (i) the impact of
PSF/diffusion kernel blurring of the cluster profile is more
important, and (ii) the fit also includes the wings, which are more
extended for non-Gaussian than for Gaussian profiles with identical
FWHM. Therefore, we expect the sizes of small clusters to be
overestimated.

\item The 5-pixel boundary adopted was estimated from the size and 
shape of the cluster light profile alone. The application of PSF and
diffusion kernel do not only change the size, but also the shape of
the cluster profile. This causes unpredictable shifts in this
empirical boundary. Nevertheless, we emphasise that the fit residuals
are very small, as we will show below.
\end{itemize}

Since King models with large concentration indices and EFF models with
small power-law index deviate most significantly from Gaussian
profiles, we expect the largest deviations from a one-to-one relation
between input and output FWHM for such models.

\subsubsection{King profiles} 

\begin{figure}
    \begin{center}
    \includegraphics[angle=180,width=0.8\columnwidth]{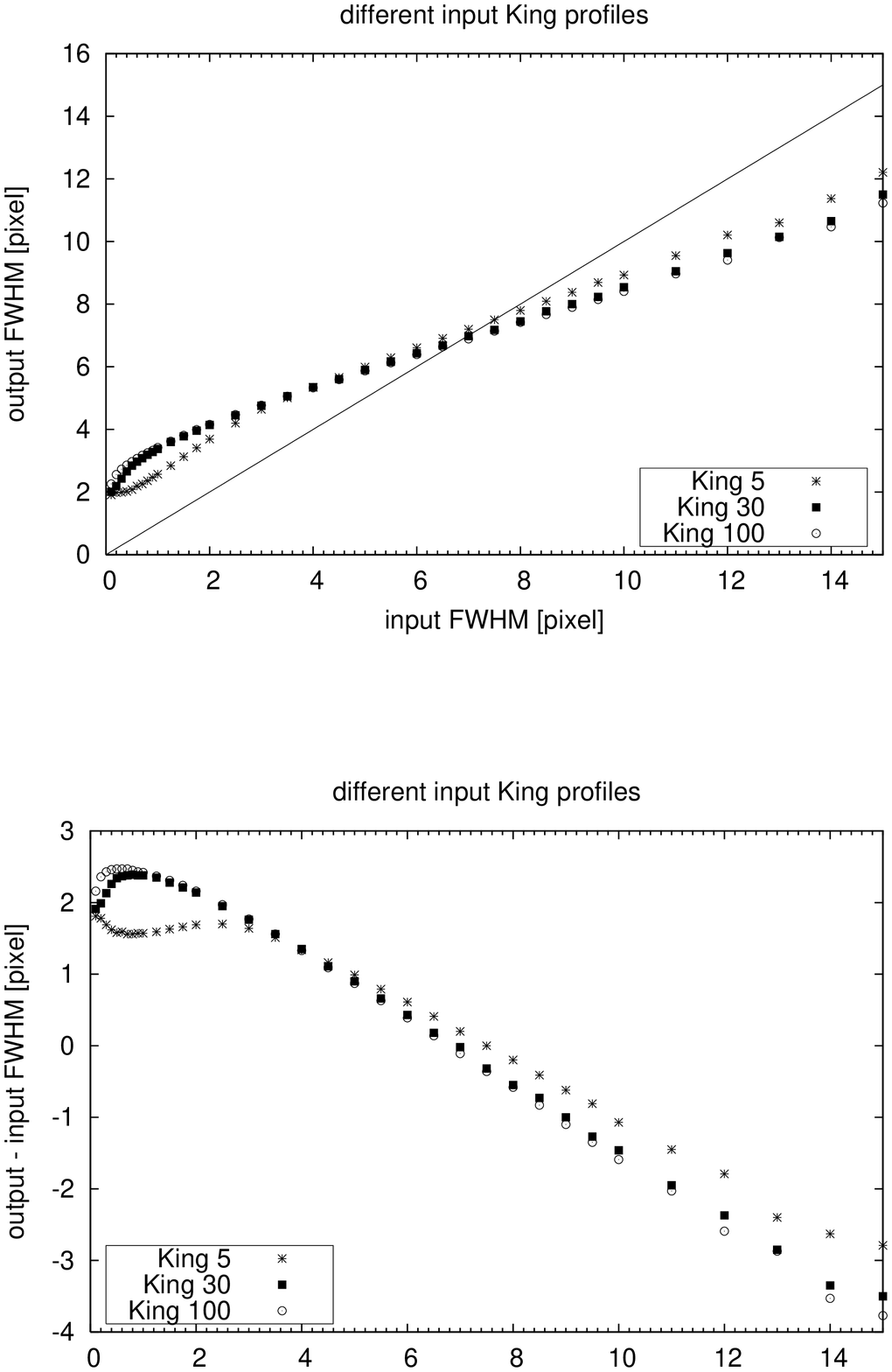}
    \end{center} \vspace{0.6cm} \caption{Fitted Gaussian FWHM for
    input King profiles with different concentrations: $\ast$, $c$ =
    5; $\blacksquare$, $c$ = 30; $\odot$, $c$ = 100. The diagonal solid
    line represents a one-to-one relation. }  \label{fig:size_king}
    \begin{center}
    \includegraphics[angle=180,width=0.75\columnwidth]{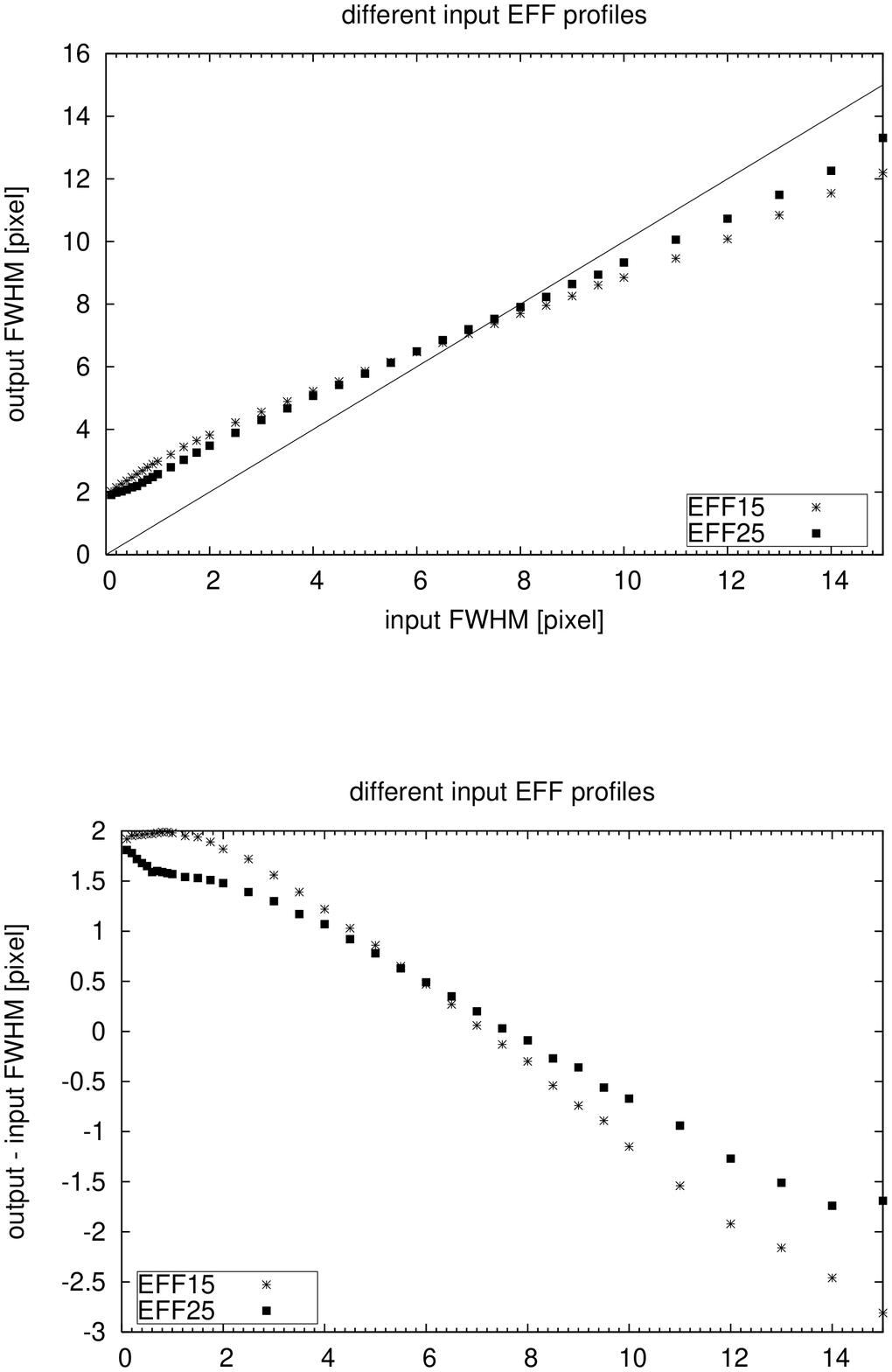}
    \end{center} \vspace{0.6cm} \caption{Fitted Gaussian FWHM for
    input EFF profiles with different power-law indices: $\ast$,
    $\gamma$ = 1.5; $\blacksquare$, $\gamma$ = 2.5.  The diagnonal solid
    line represents a one-to-one relation. }  \label{fig:size_EFF}
\end{figure}

For King profiles the results of this exercise are shown in Fig.
\ref{fig:size_king}. As expected, we find that for the less
concentrated King-profile clusters, the relation between input and
output FWHM deviates most from the relation for a Gaussian input
model, i.e., from a strict one-to-one relation. The differences
between King 5 and King 100 profiles reach $\sim 1$ pixel, with the
King 5 results lying closer to the one-to-one relation.

\subsubsection{EFF profiles}

For young clusters in the LMC, which do not show any signs of tidal
truncation, the best fit to the light distribution is a power law
(Elson et al. 1987).

Fig.~\ref{fig:size_EFF} shows the relation between input EFF-model
FWHM and the FWHM of the Gaussian fit. The same systematic
underestimate of large cluster sizes using Gaussian fits is observed
as for the King models in the previous section, as expected. The
differences between EFF 15 and EFF 25 profiles reach $\sim 1$ pixel,
with the EFF 25 profiles lying closer to the one-to-one relation.

\subsubsection{Fitting using the respective input profiles}
\begin{figure}
    \begin{center}
    \includegraphics[angle=180,width=0.75\columnwidth]{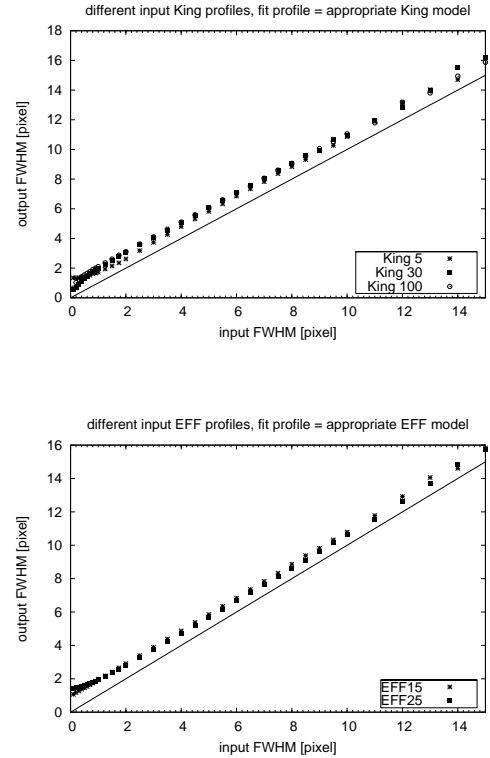}
    \end{center} \vspace{0.6cm} \caption{Fitting standard artificial clusters
    with different input profiles, using the input profile shape as fitting
    profiles (instead of the Gaussian). The diagnonal solid
    line represents a one-to-one relation. Top: Different King profiles.
    Bottom: Different EFF profiles.}  \label{fig:size_profilefit}
\end{figure}

To disentangle the effects of assuming a Gaussian profile (instead of the 
assumed input profile) on one hand and of the PSF/diffusion kernel
on the other we ran a set of simulations using the input profile
as fitting profile (instead of a Gaussian). The results presented in 
Fig. \ref{fig:size_profilefit} indicate that the strong non-linearity seen 
in Figs. \ref{fig:size_king} and \ref{fig:size_EFF} originate from using the Gaussian
fitting profile instead of the ``correct'' (input) profile. Using the input profile 
as fitting profile causes only a general offset (broadening of the
light profile due to the PSF and the diffusion kernel). Unfortunately, the 
use of EFF/King profiles for light profile fitting is not implemented as 
standard even in some uptodate image reduction software packages, while a 
Gaussian is (to our best knowledge). Therefore, while sticking to the generally
applicable Gaussian fitting profile we point at the origin of the non-linearities 
of our results.

\subsubsection{Presentation of the fit results}

We fit the relation between the input FWHM of various profiles and the
output FWHM of the Gaussian fits using a fifth-order polynomial
function. One example table for the conversions that relate the input to the
output FWHM, and vice versa, is presented in Appendix
\ref{app:size_par}, Table \ref{tab:size_standard}. 
The latter relation is most important to deduce
the intrinsic size of a source from the measured size.

For the size-dependent aperture corrections (which will be determined 
in Sect. \ref{sec:ac}) two example tables are included: one table for 
aperture corrections to infinite aperture as a function of the {\sl intrinsic}
FWHM of the object (presented in Appendix \ref{app:ac_intr}, Table 
\ref{tab:ac_wf3_inf_intr}), and one table for aperture corrections to 
infinite aperture as a function of the {\sl measured} FWHM of the object 
(presented in Appendix \ref{app:ac_meas}, Table \ref{tab:ac_wf3_inf_meas}).

In the following subsections, we will show the results for the two
physically most interesting input models only, the King 30 and the
EFF 15 models. These represent the average cluster light profiles of
old Milky Way GCs (e.g., Binney \& Tremaine 1998) and YSCs in the LMC
(Elson et al. 1987), respectively. Although the realistic light
profiles differ significantly from a Gaussian profile, Figs.
\ref{fig:size_king} and \ref{fig:size_EFF} show that the fits to our
conversion relations are very accurate. For input FWHM $>$ 0.5 pixel
the deviations of the fits from the data are always smaller than 4 per
cent, while for smaller FWHM it might be as large as 10 per cent. The
conversion functions for the standard cluster and for the full set of 
input models are given in Table~\ref{tab:size_standard}.

All data are also available in electronic form from our website, at\\
http://www.astro.physik.uni-goettingen.de/$\sim$galev/\\panders/Sizes\_AC/ .
This public dataset does not only include the parameters of the fitted 
conversion functions, but also the averaged data used for the fitting to
allow for customized fit functions, interpolations etc.

\subsection{Effect of cluster brightness: Fits and fit errors} 
\label{sec:size_mag_error}

Thus far, we considered bright, noiseless artificial clusters of a
given magnitude ($V=10$ mag), ``observed'' in a 1s exposure.

\begin{figure}
    \begin{center}
    \includegraphics[angle=180,width=0.75\columnwidth]{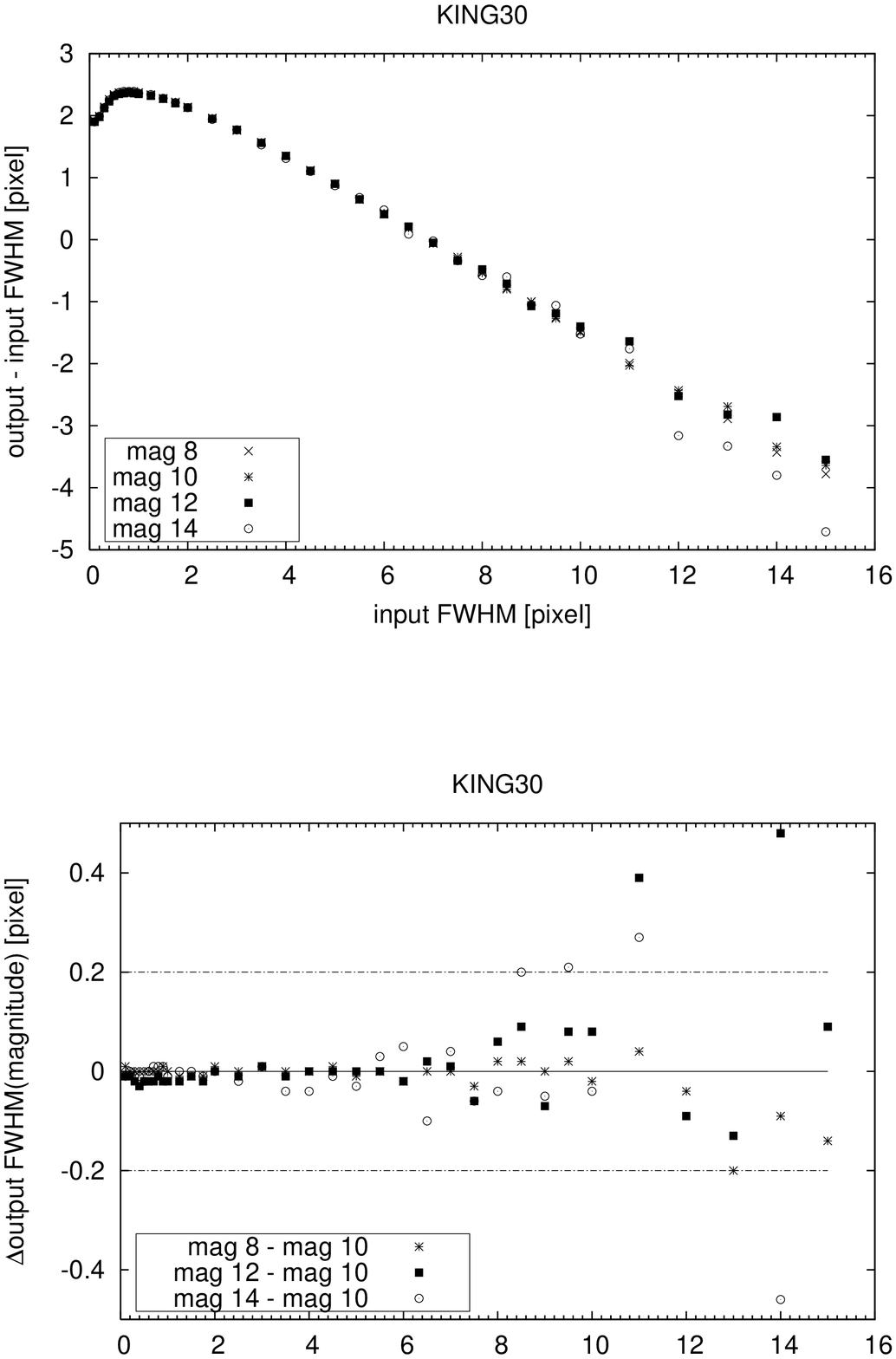}
    \end{center} 
    \vspace{0.6cm} 
    \caption{Fitted Gaussian FWHM for
    input King 30 profiles of different cluster magnitudes. Top: Output
    $-$ input sizes, simulated data and fitted polynomials. Bottom:
    Comparison of fit functions, using the $V=10$ mag fit function as
    reference.}  \label{fig:size_kingmag} 
    \begin{center}
    \includegraphics[angle=180,width=0.75\columnwidth]{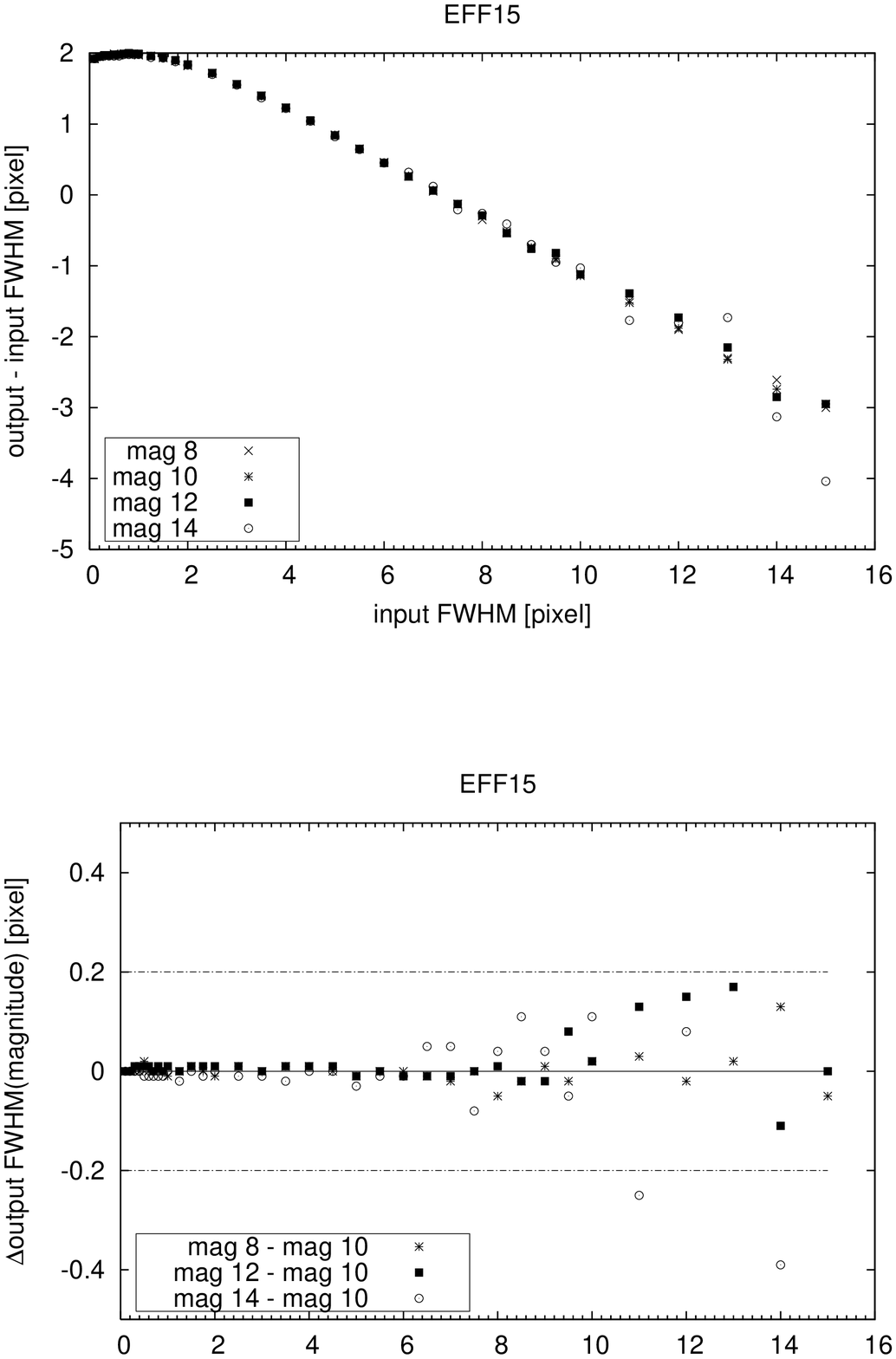}
    \end{center} 
    \vspace{0.6cm} 
    \caption{Fitted Gaussian FWHM for
    input EFF 15 profiles of different cluster magnitudes. Top: Output
    $-$ input sizes, simulated data and fitted polynomials. Bottom:
    Comparison of fit functions, using the $V=10$ mag fit function as
    reference.}  \label{fig:size_EFFmag}
\end{figure}

\begin{figure}
    \begin{center}
    \includegraphics[angle=180,width=0.75\columnwidth]{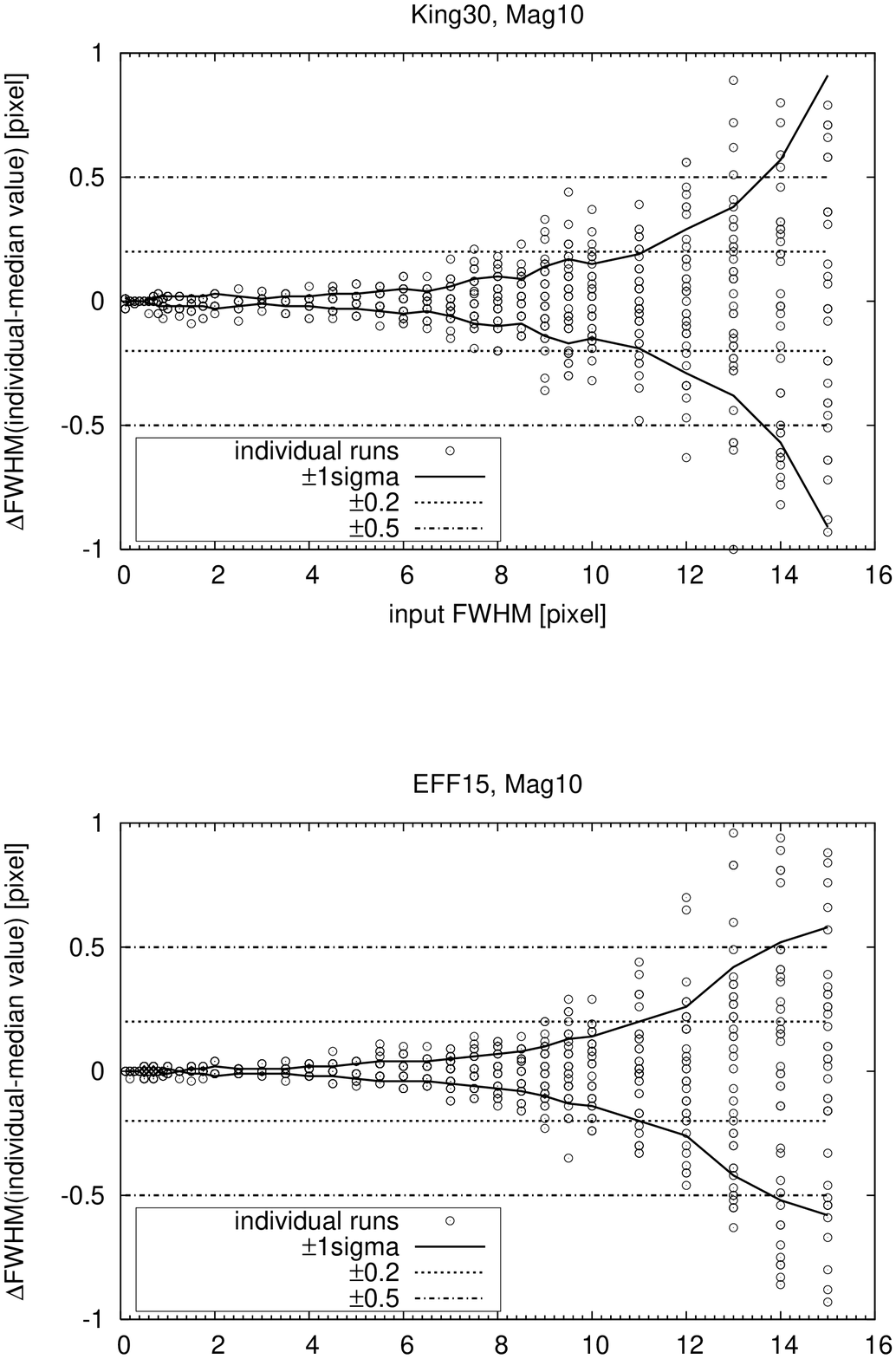}
    \end{center} 
    \vspace{0.6cm}    
    \caption{Comparison for $V=10$ mag King 30 (upper panel) and
    EFF 15 (lower panel) clusters. Horizontal lines indicate $\pm$ 0.2
    pixels and $\pm$ 0.5 pixels. The solid curved lines indicate the
    $\pm$1$\sigma$ range of the scatter. Shown are the differences of
    the individual runs with respect to the average value.}  
    \label{fig:size_seedK}
    \begin{center}
    \includegraphics[angle=180,width=0.75\columnwidth]{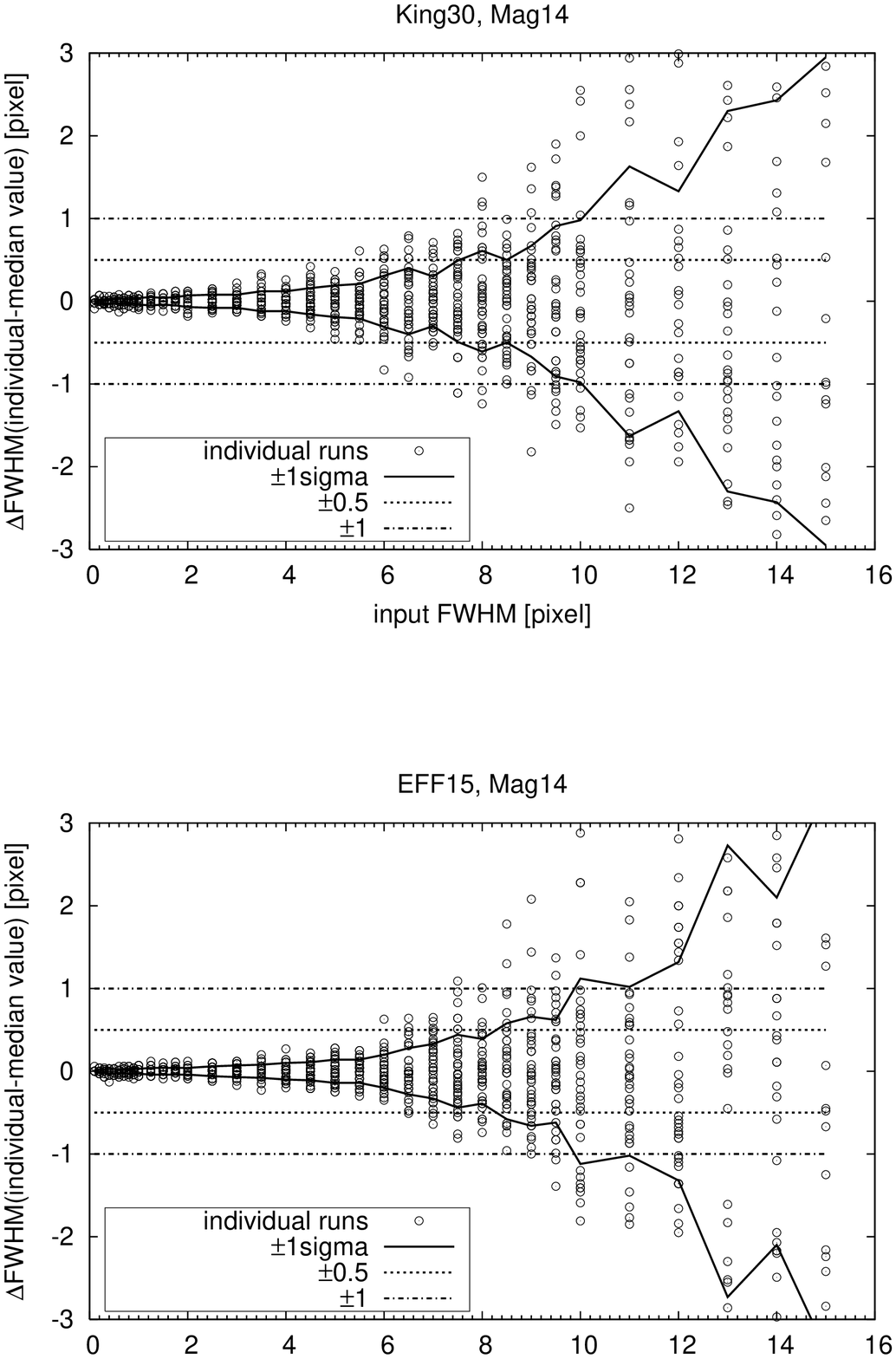}
    \end{center} 
    \vspace{0.6cm}
    \caption{Comparison for $V=14$ mag King 30 (upper panel) and
    EFF 15 (lower panel) clusters. Horizontal lines indicate $\pm$ 0.5
    and $\pm$ 1.0 pixels. The solid curved lines indicate the
    $\pm$1$\sigma$ range of the scatter. Shown are the differences of
    the individual runs with respect to the average value.}  
    \label{fig:size_seedE}
\end{figure}

We performed a series of simulations, varying the cluster magnitudes
from $V=8$ mag to $V=14$ mag. The results are shown in Figs.
\ref{fig:size_kingmag} and \ref{fig:size_EFFmag}. For each magnitude
and cluster profile, the results from 40 independent runs were averaged
to reduce the scatter. The data from the individual runs, including the 
associated 1$\sigma$ uncertainties are compiled
in Figs.~\ref{fig:size_seedK} and \ref{fig:size_seedE} to show the
amount of scatter.

Clearly, the conversion relations depend only weakly on the magnitude
of the cluster. Deviations arise because the scatter in the relation
increases with decreasing S/N ratio per pixel, as e.g. caused by decreasing
cluster brightness and/or increasing cluster sizes. In addition, for clusters 
with sauch low S/N ratios per pixel, the readout noise might have some impact. See Sect. 
\ref{sec:size_sky} for further details.

When we compare the results from Figs.~\ref{fig:size_kingmag} and
\ref{fig:size_EFFmag} to those of Figs.~\ref{fig:size_seedK} and
\ref{fig:size_seedE}, we see that the scatter/deviations in the former
figures for fainter clusters is fully consistent with the intrinsic
scatter caused by the random nature of the algorithm used to generate
the artificial clusters, even for averaged results. For bright
clusters (represented by the $V=10$ mag cluster), the random scatter
is on the order of $\pm 0.1$ pixel in the FWHM (for clusters larger
than about 8 pixels FWHM up to $\pm 0.2$ pixel in the FWHM, and up to
$\pm 1$ pixel for clusters larger than 11 pixels FWHM). This scatter
increases with decreasing cluster brightness (up to $\pm 0.4-0.5$
pixels in the FWHM for a $V=14$ mag cluster smaller than about 7
pixels FWHM, and up to $\pm 3$ pixels in the FWHM for clusters larger
than 7 pixels FWHM). The increasing scatter for large clusters is
caused by the lower S/N ratio per pixel. The data are all presented in
terms of the individual absolute values from the different runs to
illustrate the scatter. The median value scatters significantly less.

{\bf In summary, for the average cluster, the conversion relations for bright
clusters can be applied to clusters of all magnitudes: For fainter clusters and
at larger radii (hence for cases with low S/N ratios per pixel) the cluster-to-cluster
variations get larger, but scatter symmetrically around the average conversion
relation.}

\subsection{Fitting radius variations}
\label{sec:size_fitrad}

\begin{figure}
   \begin{center}
   \includegraphics[angle=180,width=0.75\columnwidth]{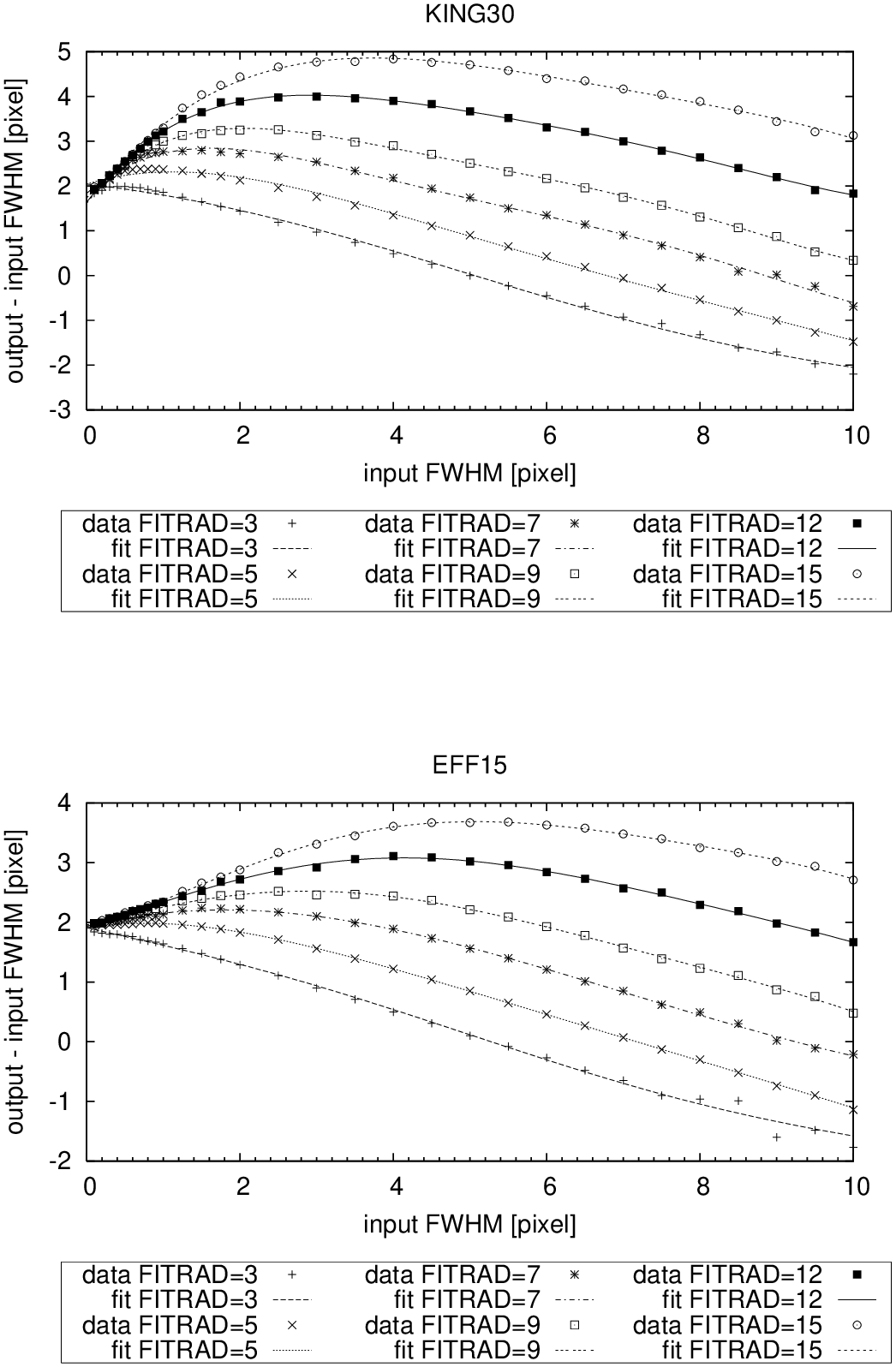}
   \end{center} 
   \vspace{0.4cm} 
   \caption{Conversion relations for a
   standard cluster, using different fitting radii (given in
   pixel units in the legend).}  
   \label{fig:size_fitraddata} 
   \begin{center}
   \includegraphics[angle=180,width=0.75\columnwidth]{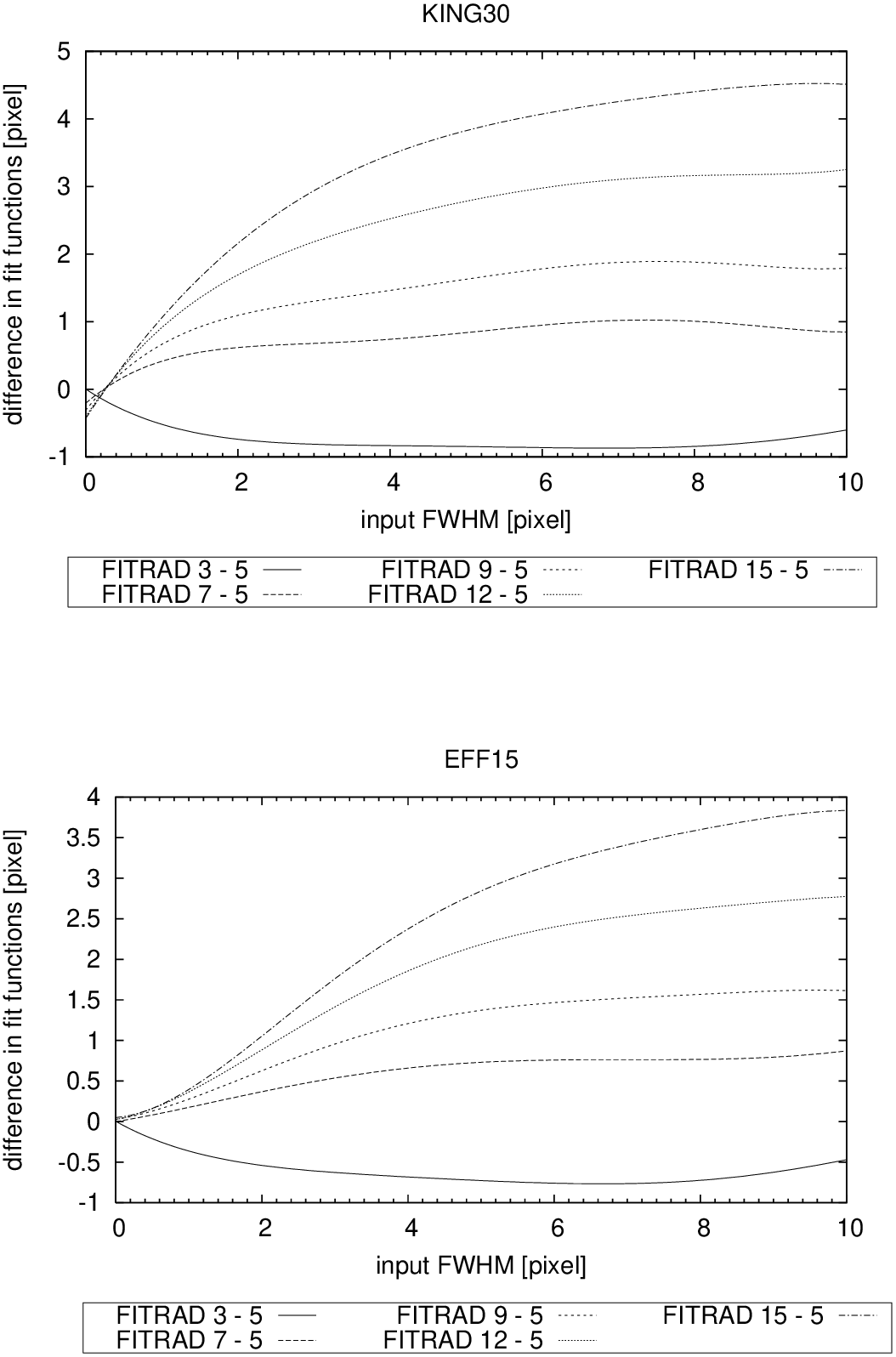}
   \end{center} 
   \vspace{0.6cm}
   \caption{Conversion relations for a standard
   cluster, using different fitting radii (given in pixel units in the
   legend). Displayed are the offsets of these relations from the
   relation for a fitting radius of 5 pixels.}  
   \label{fig:size_fitrad}
\end{figure}

{\sc BAOlab} has the advantage that the fitting radius can be
adjusted easily. In fact, the choice of fitting radius has a major
impact on the cluster sizes that one determines, as we will show
in this section. We performed tests using fitting radii in the
range from 3 to 15 pixels (larger and smaller fitting radii did
not lead to any meaningful results owing to numerical problems
related to the convergence of the size fitting). As one can see
from Figs. \ref{fig:size_fitraddata} and \ref{fig:size_fitrad},
the larger the fitting radius one adopts, the larger the apparent
cluster radius one measures, and the stronger the deviations from
the input values become. In fact, increasing the fitting radius
seems to result in continuously increasing recovered cluster
radii. This is caused by the impact of (i) the intrinsic profile
mismatch between King and EFF profiles, and (ii) the
PSFs/diffusion kernels and their non-Gaussianity. The fitting
radius dependencies of the results will be significantly lower if
one were to fit the clusters with the correct cluster light
profile, including the right PSF and diffusion kernel. However,
since we wanted to keep our study as generally applicable as
possible, we did not make use of the respective functions {\sc
BAOlab} provides in the standard settings. However, we refer the
reader to Section \ref{sec:size_PSFfit}, where we discuss this in
more detail.

As shown by Carlson \& Holtzman (2001), even fitting King profiles
(which are thought to be more realistic, at least for old globular
clusters) to observed cluster profiles is fitting radius dependent.
They attribute this behaviour to inaccuracies of the PSFs at small
radii. And indeed, their situation is different from ours, in the
sense that in our case we expect the intrinsic differences of the
input profiles and the fitted Gaussian to dominate the fitting
behaviour, not inaccuracies of the PSFs, while for Carlson \& Holtzman
(2001) the profile mismatch, if any, is likely smaller.

A selection of fit residuals is included in Appendix
\ref{app:illustration}, in Figs.~\ref{fig:9king} and \ref{fig:9eff},
as a function of fitting radius and input cluster radius. The area
shown covers the inner $5\times5$ pixels. For fitting radii $<5$
pixels, the solution tends to become computationally unstable, as
shown in Fig.~\ref{fig:size_fitraddata}.

For small clusters, the residuals shown in Figs.~\ref{fig:9king} and
\ref{fig:9eff} are almost independent of the fitting radius, because
in all cases the cluster is much smaller than the fitting radius.
However, the residuals are significantly non-negligible, clearly
showing the intrinsic difference in shape between Gaussian and
King/EFF profiles.

For large clusters (we show the results for clusters with FWHMs of 5.0
and 10.0 pixels, respectively), the residuals are relatively small for
small fitting radii (e.g., fitting radii on the order of the input
FWHM), where the fit is dominated by the inner parts of the clusters,
which resemble Gaussian profiles. For fitting radii greater than the
FWHM, the cluster wings are given too much weight, resulting in strong
deviations in the inner cluster parts and large residuals (just as for
``small'' clusters, discussed above). The maximum residuals increase
by a factor 3--5 for fitting radii from 5 to 15 pixels. However,
fitting the inner cluster parts only seems to be more promising for 2
reasons: (i) the inner cluster region resembles a Gaussian profile
more closely, and hence fitting with a Gaussian function is less
problematic, and (ii) the S/N ratio per pixel is higher in the inner
parts than in the wings.

In summary, one would like to have a fitting radius large enough to
give stable results (larger than 3 pixels, cf. Figs.
\ref{fig:size_fitraddata} and \ref{fig:size_fitrad}), but small
enough to fit mainly the cluster core rather than the wings, to
avoid serious problems with structures in the immediate environment
of the cluster (e.g., variable background, crowding effects, etc.) 
and to produce (close to) negligible deviations from the input size.
In addition, as shown in Figs.  \ref{fig:size_fitraddata} and
\ref{fig:size_fitrad}, the impact of changing the fitting radius is
such systematic and significant, that a single, generic value for
the fitting radius is needed. Otherwise, the entire analysis in this
paper must be done for each individual data set.

We therefore recommend the use of a generic fitting radius of 5
pixels (which should be applicable to almost all realistic
observations), and emphasise that all results given in this paper
were thus obtained.

\subsubsection{Origin of the strong fitting radius dependence of the
results} 

The cause of the strong fitting radius dependence of our size
conversion relations most likely also causes the non-linearity of the
size conversion relations, i.e., the shape difference between the
intrinsic cluster profile (EFF or King profiles) and the Gaussian used
for the fitting.

\begin{figure}
   \begin{center}
   \includegraphics[angle=180,width=0.75\columnwidth]{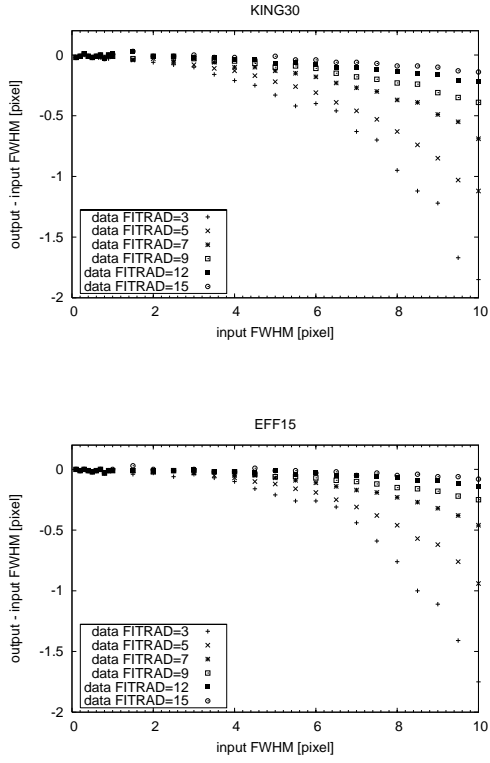}
   \end{center} 
   \vspace{0.6cm}
   \caption{Conversion relations for a standard
   cluster, using different fitting radii (given in pixel units in the
   legend) and taking the input light profile, PSF and diffusion kernel into account.}  
   \label{fig:size_fitrad2}
\end{figure}

To test this hypothesis we have performed a set of simulations similar
to the ones in the previous section, except now the fit models are the
same as the input models, and they were convolved internally with the
appropriate PSF and the diffusion kernel. The results, shown in Fig.
\ref{fig:size_fitrad2}, partially support our hypothesis, even though 
for large fitting and cluster radii the behaviour is still non-linear, 
and differs systematically among the fitting radii.

We conclude that in order to get a one-to-one correlation between input and 
output FWHM the fitting radius must be at least larger than the cluster radius.
In case the fitting radius equals the cluster radius, the deviations from a 
one-to-one correlation are typically of the order of -0.2/-0.3 pixel, as can be 
seen in Fig. \ref{fig:size_fitrad2}. However, these deviations/non-linearities
are intrinsically taken into account in our results.

\subsection{Impact of the sky background}
\label{sec:size_sky}

\begin{figure}
   \begin{center}
   \includegraphics[angle=180,width=0.75\columnwidth]{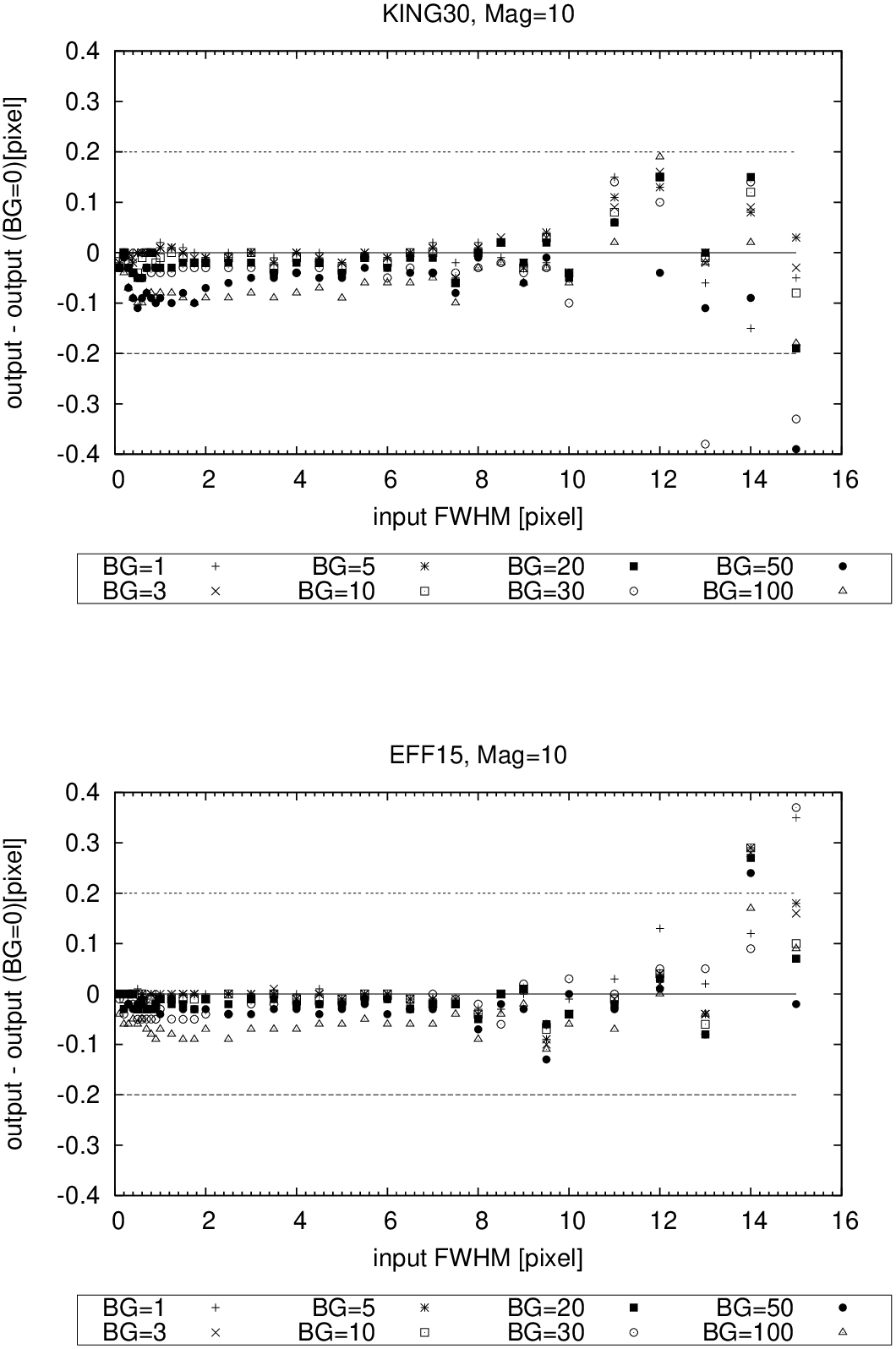}
   \end{center} \vspace{0.4cm} \caption{Conversion relations for a
   $V=10$ mag cluster, taking sky noise into account. As reference, the
   data for a standard cluster are taken. The straight lines are at $\pm$ 0.2 pixel.}
   \label{fig:size_skymag10} 
   \begin{center}
   \includegraphics[angle=180,width=0.75\columnwidth]{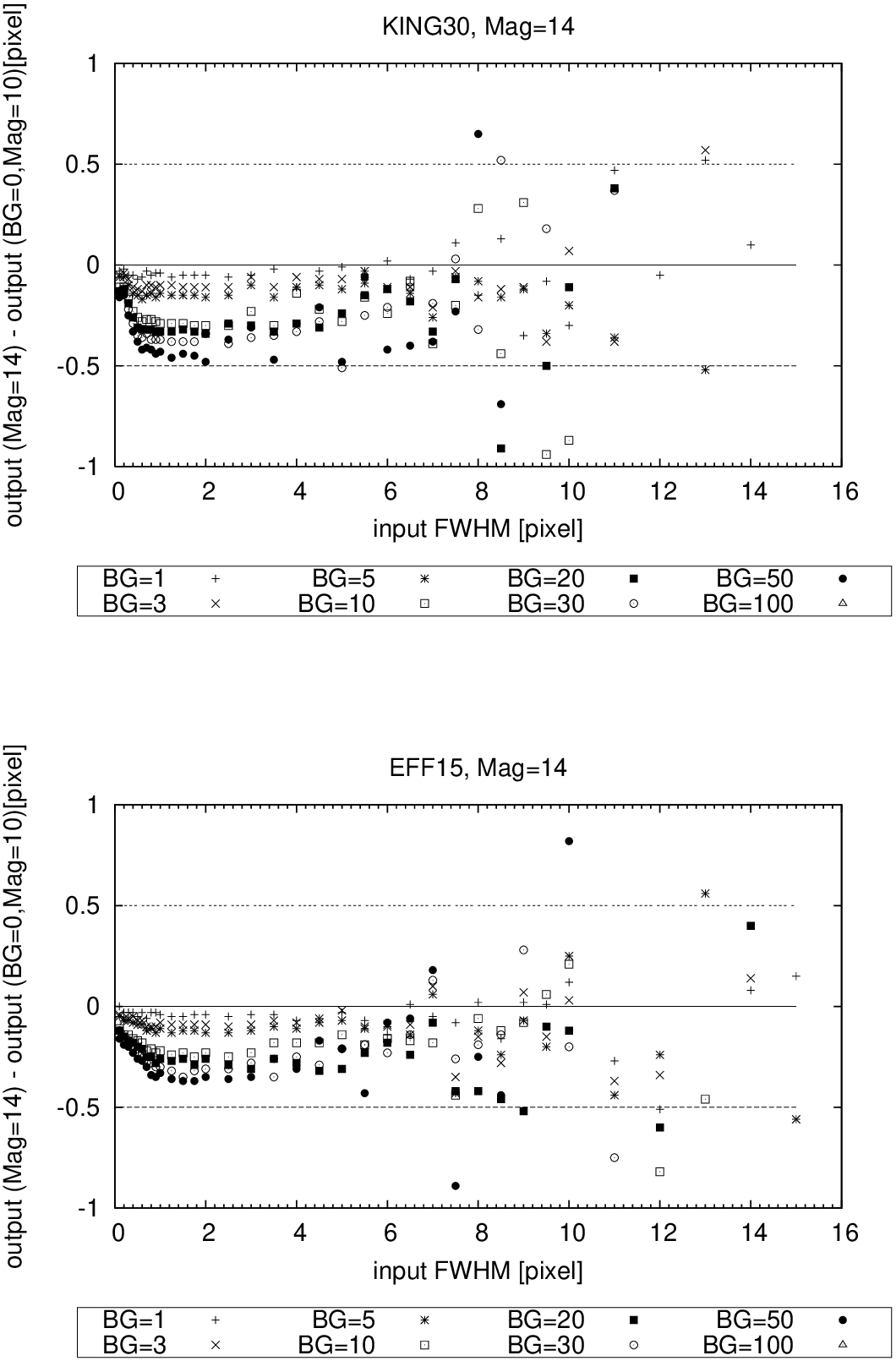}
   \end{center} \vspace{0.4cm} \caption{Conversion relations for a
   $V=14$ mag cluster, taking sky noise into account. As reference,
   the data for a standard cluster are taken. The straight lines are at $\pm$ 0.5 pixel.}
   \label{fig:size_skymag14}
\end{figure}

\begin{figure}
   \begin{center}
   \includegraphics[width=0.75\columnwidth]{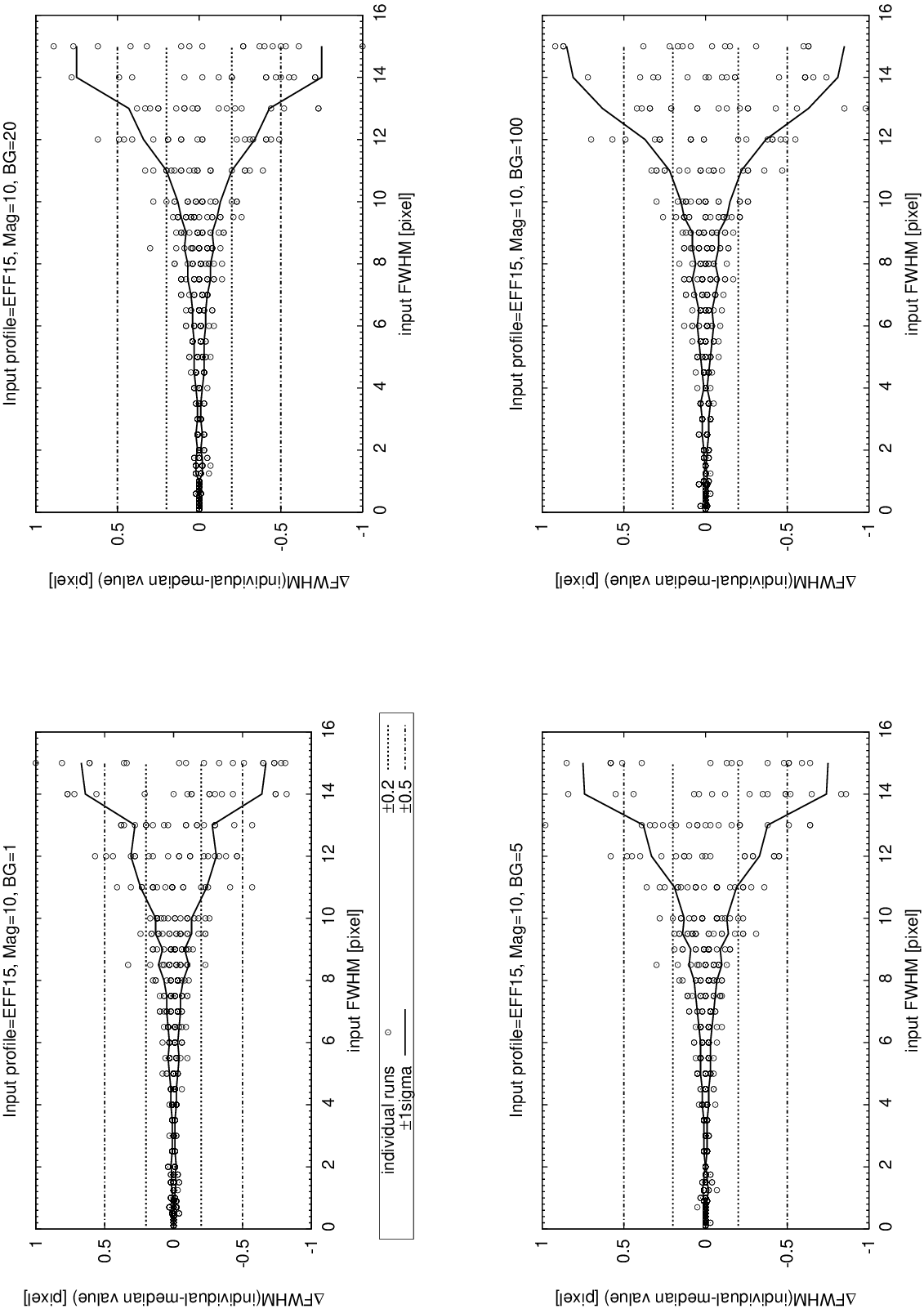}
   \end{center} \vspace{0.4cm} \caption{Scatter for $V=10$ mag EFF 15
   clusters, with varying sky levels (``BG'', in ADU). As
   reference, the average data are taken. The straight lines are at
   $\pm$ 0.2 and $\pm$ 0.5 pixels. The solid curved lines indicate the
    $\pm$1$\sigma$ range of the scatter.}  
   \label{fig:size_skyScatter10}
   \begin{center}
   \includegraphics[width=0.75\columnwidth]{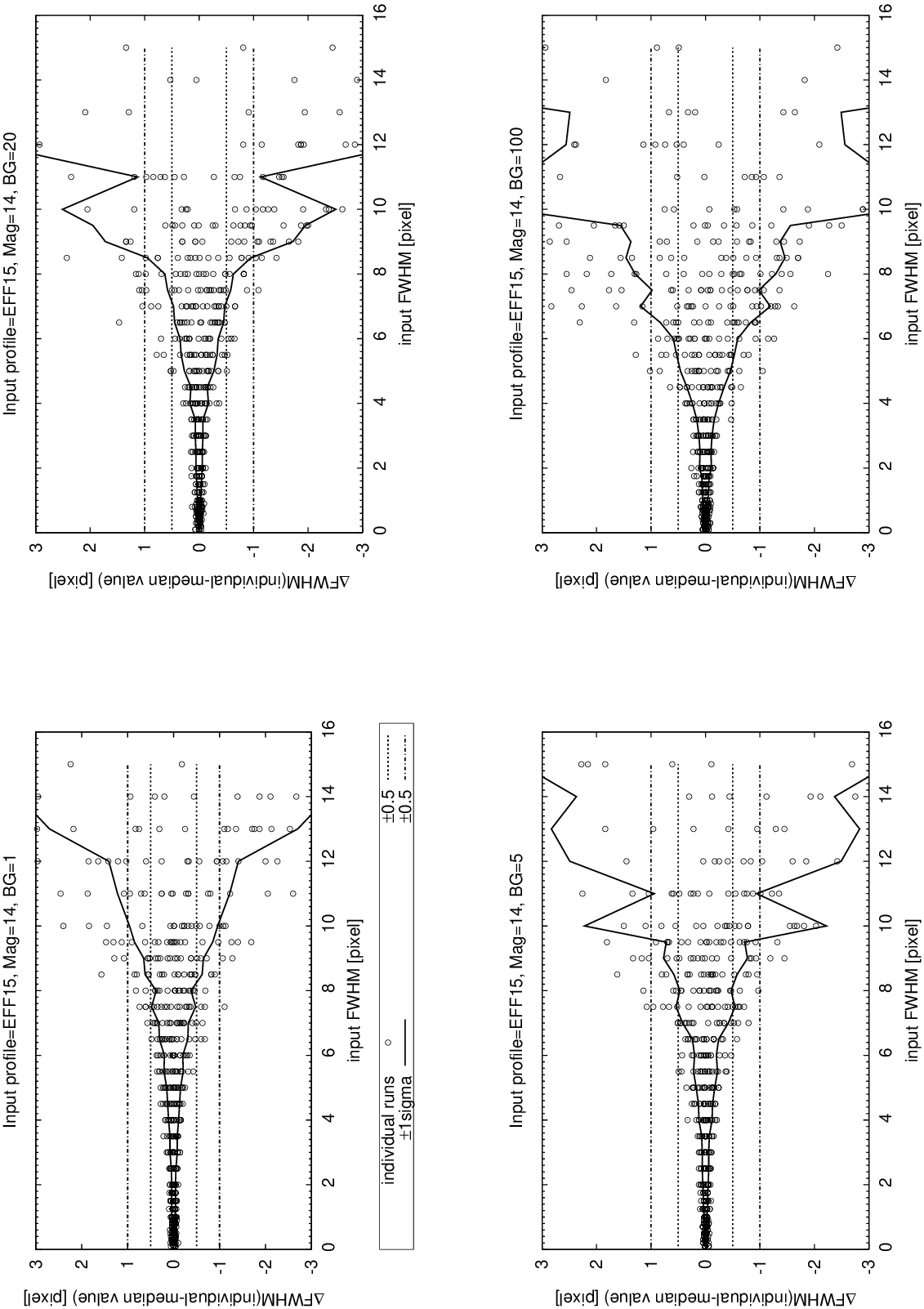}
   \end{center} \vspace{0.4cm} \caption{Scatter for $V=14$ mag EFF 15
   clusters, with varying sky level (``BG'', in ADU). As
   reference, the average data are taken. The straight lines are at
   $\pm$ 0.5 and $\pm$ 1.0 pixels. The solid curved lines indicate the
    $\pm$1$\sigma$ range of the scatter.}
   \label{fig:size_skyScatter14}
\end{figure}

In this section we assess the importance of the sky background. We
model $V=10$ mag and $V=14$ mag clusters, each with background levels of 0, 1, 3,
5, 10, 20, 30, 50, and 100 ADU (``counts'') per pixel. We use 20
independent runs, and show the averaged results in Figs.
\ref{fig:size_skymag10} and \ref{fig:size_skymag14}, and the
associated plots illustrating the actual scatter in Figs.
\ref{fig:size_skyScatter10} and \ref{fig:size_skyScatter14} for the
selected magnitudes, $V=10$ and $V=14$ mag, respectively.

Since sky noise and readout noise have the same characteristics, this Section
combines both effects.

On average, the results seem to be robust with respect to (constant)
changes in the sky background. There is only a slight tendency for
faint clusters on a strong sky background to appear marginally smaller
(see Fig. \ref{fig:size_skymag14}).

The impact of (Poissonian) shot noise from the cluster itself is
negligible.

In order to allow for a {\bf rough} estimate of the S/N ratios for the clusters
and background levels discussed in this Section we provide the approximate 
count rates in the peak pixel of selected clusters in Table \ref{tab:counts}.

\begin{table}[!h]
\caption{Count rates [in ADU] in the peak pixel of selected clusters, 
to approximate S/N ratios for the background levels}
\begin{center}
\begin{tabular}{@{}*{5}{|c}{|}@{}}
\hline
cluster's FWHM & King 30 & EFF 15  & King 30 & EFF 15 \\
$[$pixel$]$    & $V=10$ & $V=10$ & $V=14$ & $V=14$\\
\hline
1.0   & 6200 & 7600 & 150 & 180 \\
2.0   & 3100 & 4100 & 80  & 100 \\
5.0   & 900  & 1200 & 25  & 30  \\
10.0  & 300  & 400  & 8   & 10  \\
\hline
\end{tabular}
\label{tab:counts}
\end{center}
\end{table}

\subsection{Using the appropriate PSFs for fitting}
\label{sec:size_PSFfit}

For four {\sl HST}/WFPC2 filters we checked to what extent
the filter-dependent PSFs affect the cluster size determinations.
{\sc BAOlab} allows one to consider the appropriate PSF when
fitting the cluster size by convolving the Gaussian model clusters
with the PSF specified. Hence we created clusters with PSFs for
the {\sl HST}/WFPC2 $U, B, V, I$-band equivalent filters; while
fitting the size of the cluster, {\sc BAOlab} took the appropriate
PSF into account. Here, we investigate the impact and
possibilities of this approach.

\begin{figure}
   \begin{center}
   \includegraphics[angle=180,width=0.75\columnwidth]{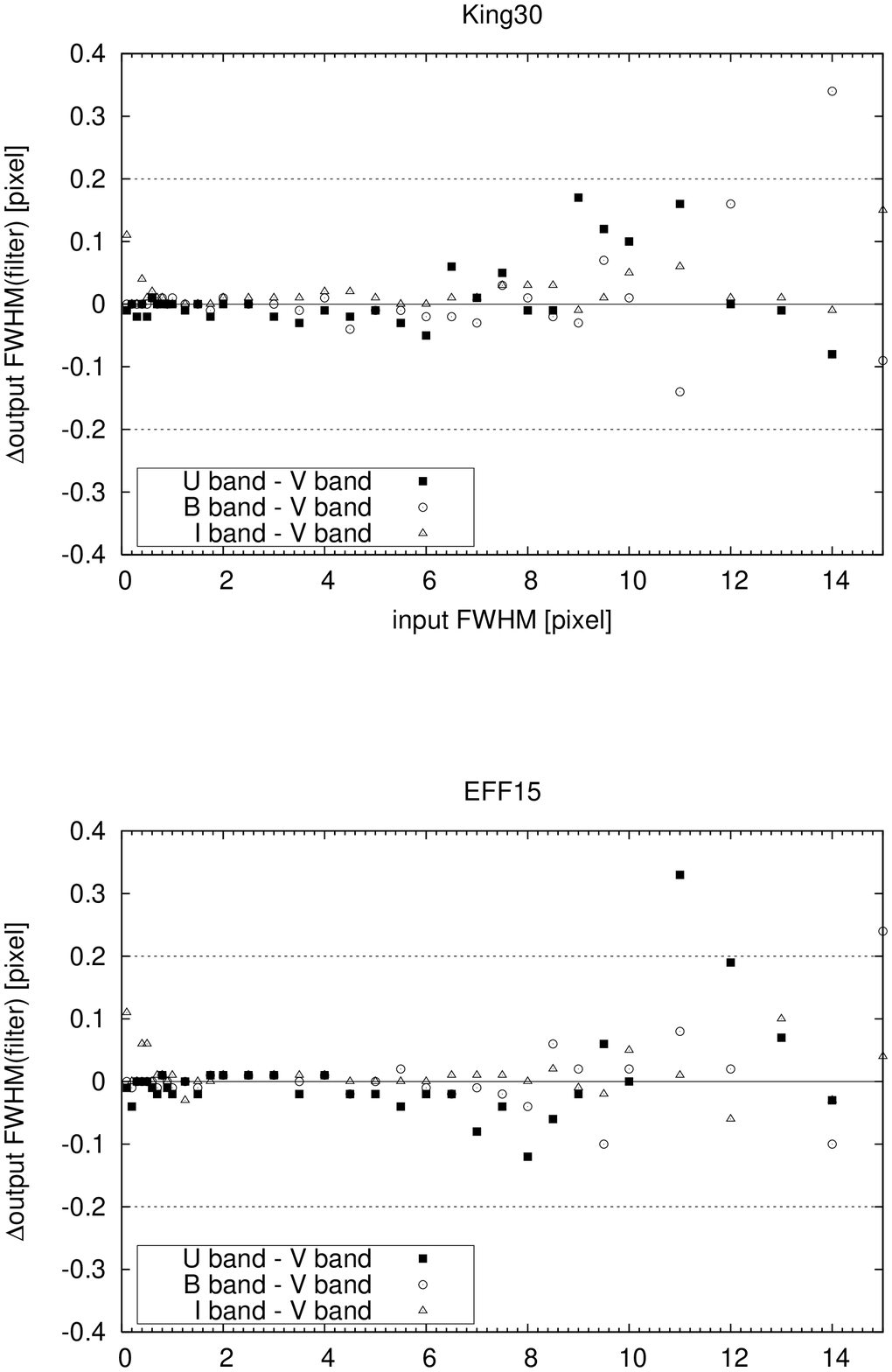} 
   \end{center}
   \vspace{0.4cm}
   \caption{Conversion relations for a standard cluster, using the
   appropriate PSFs for the fitting. The horizontal lines indicate $\pm$0.2 pixel.}  
   \label{fig:size_right_PSF} 
   \begin{center} 
   \vspace{0.4cm}
   \includegraphics[angle=180,width=0.75\columnwidth]{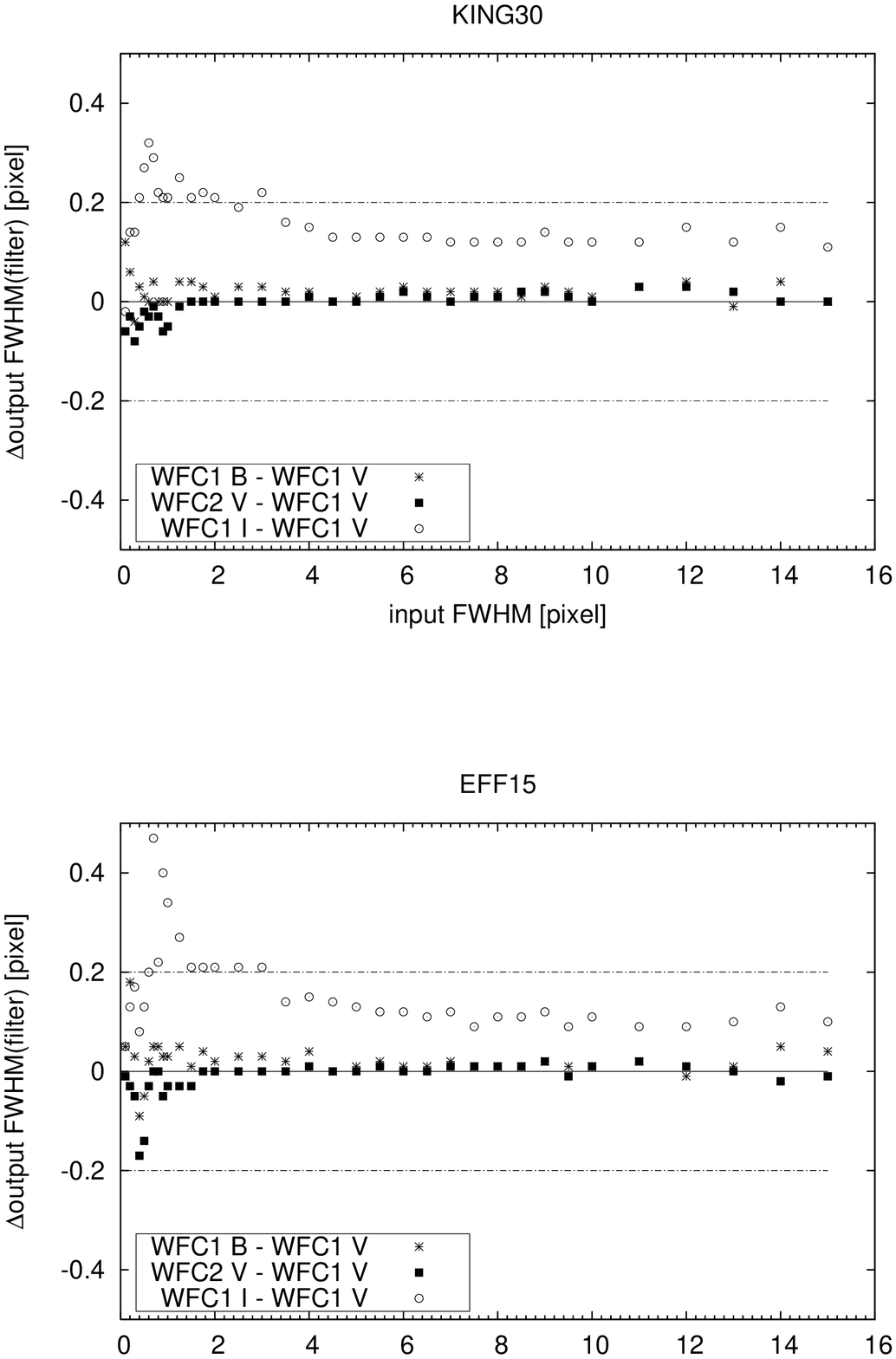} 
   \end{center} 
   \vspace{0.4cm} 
   \caption{Conversion relations for a standard cluster, using different filters for the ACS/WFC (chip 1).}   
   \label{fig:size_ACSFilter} 
\end{figure}

The results are presented in Fig.~\ref{fig:size_right_PSF}. While the
relations appear to be slightly noisier (despite the use of 40
independent runs), the differences between the different filters are
well within the ``normal'' scatter of $\pm 0.2$ pixel.

\subsection{Other dependences}
\label{sec:size_filter}

We investigated the results for different WFPC2 filters and chips.
Both variations lead to only minor differences in the results for
the ``standard'' cluster. However, for completeness we give the
results at our webpage.

We also checked the dependence of the results on the position of
the artificial cluster on the chip, and on subpixel shifts. Both
tests gave results within the ``standard'' random scatter of $\pm
0.2$ pixel.

In addition, we produced a number of PSFs assuming various
spectral types for the standard cluster, ranging from O5 to M3.
Again, all differences remain within the usual random scatter of
$\pm 0.2$ pixel.

These ``non-dependencies'' are consistent with Carlson \& Holtzman
(2001), who found that different PSFs only have a minor impact.

\subsection{Observing with ACS: chip, position, and filter dependence}
\label{sec:size_acs}

Since both of the {\sl HST} ACS/WFC chips (WFC1 and WFC2) are located
well off the instrument's optical axis, the PSFs suffer from severe
geometrical distortions and the diffusion kernel is both wavelength
and position dependent. Therefore, we carried out simulations with
{\sc Tiny Tim} PSFs for both WFC chips, for various positions on the
chip, and -- for the central positions of each camera -- also using
different filters (F435W, F555W, and F814W, roughly equivalent to
Bessell-Johnson-Cousins $B, V$ and $I$). Again, 40 independent runs
were used to obtain average values with reduced scatter.

The results for the different filters used with the ACS/WFC are shown
in Fig. \ref{fig:size_ACSFilter}. The strongest differences are
observed for the F814W filter, for which the sizes we determined are
systematically larger, by $\sim 0.1-0.2$ pixels, than those obtained
for the F555W filter. The reason is not quite clear, as both the PSF
and the diffusion kernel appear {\sl less} extended than the respective 
values for the F555W filter. In addition, Fig. \ref{fig:size_ACSFilter}
shows prominent discontinuities for the F814W filter around input FHWMs
of 0.7 pixel. These peaks are statistically significant, and apparent 
in almost every single run.

The differences in the F435W filter are
significantly smaller. In addition, when comparing the results for the
F555W passband for the two WFC chips, we find only small differences.
The deviations in the latter two cases are below or on the order of
$\pm 0.05$ pixel.

Despite the distortion of the chips and the corresponding changes in
the PSF with position across the chip, we find only a small impact on
the derived sizes of the cluster position on the chips. For almost all
clusters the deviations are well within $\pm 0.1$ pixel. Hence, for
almost all purposes the central PSF (and the associated diffusion
kernel) can be used.

\subsection{Observing with NICMOS: filter-dependence}
\label{sec:size_nic2}

PSF construction for NICMOS using {\sc Tiny Tim} is not
straightforward, mainly because of the off-focus setting during early
observations [i.e., before servicing mission 3B]. This caused strong
temporal PSF dependences. The results discussed here are for two
distinct observation dates. After inspection of a coarse grid of PSFs
for different observation dates, we selected {\sc Tiny Tim} PSFs for
1998 February 1 (as an example of a fairly blurred PSF) and ``after
2002 September 29'' (fully installed cryocooler phase, with only minor
PSF blurring).

\begin{figure}
\begin{center} 
\vspace{0.4cm}
\includegraphics[angle=180,width=0.75\columnwidth]{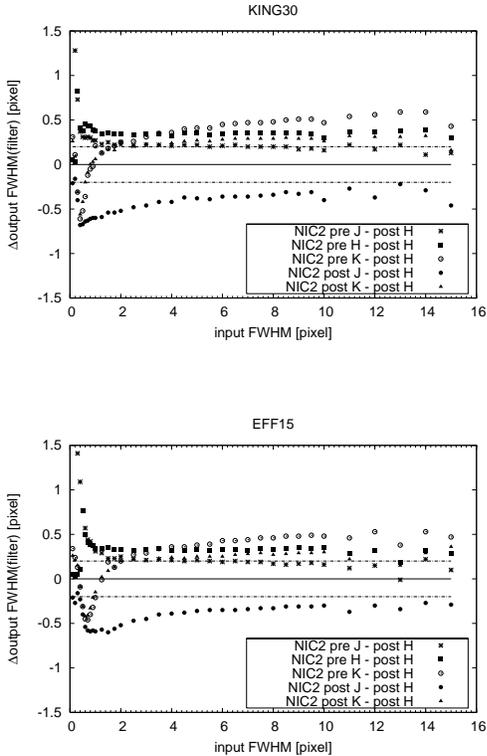} 
\end{center} 
\vspace{0.4cm} 
\caption{Conversion relations for a standard cluster (Upper panel:
King 30 profile; Lower panel: EFF 15 profile), using
different filters for NICMOS camera 2.}   
\label{fig:size_NICMOSFilter} 
\end{figure}

In both cases, a strong filter dependence is apparent, as can be seen
in Fig.~\ref{fig:size_NICMOSFilter}. For this reason, we give size
conversions for all filters analysed (NICMOS equivalents to $J$, $H$
and $K$) and both epochs of observations at our webpage.

\subsection{Further sources of uncertainties}
\label{sec:size_uncert}
Before assessing further possible sources of uncertainties we 
remind the reader that our study does NOT aim at working at the
spatial resolution limit of {\sl HST} (marginally resolved sources were 
already studied in Carlson \& Holtzman 2001), but advice the users to
apply our recipes only for clusters with an intrinsic FWHM greater than 
0.5 pixel. Hence many of the tiny details and uncertainties inherent to 
PSF modelling (both theoretically as {\sc Tiny Tim} and using observed
PSFs) smear out, becoming irrelevant for our studies.

Drizzling and jitter might blur images slightly by broadening the PSF. 
However, as Carlson \& Holtzman 2001 stated already (for an even more
difficult situation, due to their smaller clusters, compared to ours)
both effects have minor impact. The exact amount depends on the brightness
of the source, the quality of the image reduction software to perform 
subpixel alignment, the number of exposures stacked, the length of these
exposures etc..

For faint sources even at radii close to the peak the count numbers per
pixel might reach values where Poisson statistics is non-negligible. However,
this problem is intrinsic when dealing with faint sources. We have tried to
quantify this effect in Figs. \ref{fig:size_seedK}, \ref{fig:size_seedE},
\ref{fig:size_skyScatter10}, and \ref{fig:size_skyScatter14}. For a further
assessment and comparison with  the widely-used DeltaMag method see Sect.
\ref{sec:comparison}.

To fully utilize the subsampling of the PSF, the use of a subsampled 
charge diffusion kernel/subsampled response function would be best, 
including its wavelength dependence. However, these are not available, 
and not likely to become available (see {\sc Tiny Tim} FAQ page).

Other uncertainties, like breathing, desorption, scattering by dust, 
scratches or the electrode structure, the presence of ghost images etc. 
are complex and beyond any realistic measuring and modelling effort. 
However, most of these uncertainties are shared with observed PSFs. 

\section{Determining accurate photometry: Aperture corrections}
\label{sec:ac}

\subsection{Input parameters}
\label{sec:ac_input}

We generate artificial clusters of different light profiles and sizes,
$R_{\rm cl}$, using the {\sc BAOlab} package and convolve them with
the PSFs and diffusion kernels appropriate for the different cameras.
We then determine ACs as a function of the FWHM of
the source and the size of the aperture, $R_{\rm ap}$.

The aperture correction (AC$_\lambda$) is defined as:
\begin{eqnarray}
{\rm AC}_\lambda &=& -2.5\log \Bigl( F_\lambda(R_{{\rm
ref}},R_{{\rm cl}})/F_\lambda(R_{{\rm ap}},R_{{\rm cl}}) \Bigr) \cr &=& {\rm
mag}_\lambda(R_{{\rm ref}},R_{{\rm cl}}) - {\rm mag}_\lambda(R_{{\rm ap}},R_{{\rm cl}}) \quad ,
\end{eqnarray}
where $F_\lambda(R)$ is the flux within an aperture with radius $R$,
and mag$_\lambda(R)$ the corresponding magnitude, both for a given
wavelength (filter) $\lambda$. The ``ref(erence)'' radius, $R_{\rm
ref}$, is either infinity or another radius taken for reference, e.g.,
0.5 arcsec (as recommended, e.g., by Holtzman et al. 1995). However,
we will show that correcting to 0.5 arcsec, while appropriate for
point sources, is insufficient for extended objects.

We consider three cluster light profiles in the remainder of this
study:

\begin{itemize}
\item {\bf King 5}: King profiles with $c = 5$. This corresponds to
the average concentration index observed for Galactic open clusters
(e.g., Binney \& Tremaine 1998); we emphasise, however, that because
of the peculiar cluster profile and the large extent of the {\sl HST}
PSFs, the size and AC relations for King 5 profiles are more uncertain
than for the other profiles;

\item {\bf King 30}: King profiles with $c = 30$, corresponding to the
average concentration index observed for Galactic globular clusters
(e.g., Binney \& Tremaine 1998);

\item {\bf EFF 15}: Elson, Fall \& Freeman models with a power-law
index of 1.5, matching the average observed profile of young populous
LMC clusters (Elson et al. 1987).
\end{itemize}

Clusters with FWHMs between 0.1 and 15 pixels were considered, for
each chip. For the clusters with small FWHMs, PSF photometry might be
a more accurate solution. For this purpose the {\sc HSTphot} package
of Dolphin (2000) is available. However, since this package only works
with very specific data formats, we cannot present a direct comparison
of both methods here. Clusters larger than 15 pixels FWHM are very
unlikely to occur, in particular since they must be very bright to
have sufficiently high S/N ratios out to such large radii. For such
clusters, additional effects become important, including background
contamination and crowding.

The analysis was done for the WFPC2 PC and WF3 chips, for the ACS/WFC1
(WFC2 is equivalent) and for the NICMOS/NIC2 (pre and post-cryocooler)
chips.

After generating and convolving the clusters with the appropriate PSF
and diffusion kernel (where relevant), aperture photometry of the
noiseless objects was done using different apertures. The ACs
were determined with respect to the reference aperture. If
not otherwise specified $R_{\rm ref} = 50$ pixels is used. No model
exhibits strong changes in the light profile for such large radii,
hence $R_{\rm ref} = 50$ pixels is sufficiently close to $R_{\rm ref}
= \infty$ for our purposes.

\subsection{The relation between aperture correction and {\bf input} FWHM}
\label{sec:intr_AC}

First, we determine the relation between AC and
input FWHM of the object. 

The result is again fitted with a fifth-order polynomial function,
\begin{equation}
{\rm AC}(x) = a + b*x + c*x^2 + d*x^3 + e*x^4 + f*x^5 \, ,
\label{eq:ac_fit}
\end{equation}
where $x$ is the input FWHM of the object (in pixels), and $a$ through
$f$ are the fitting coefficients.

\begin{figure}
    \includegraphics[angle=180,width=\columnwidth]{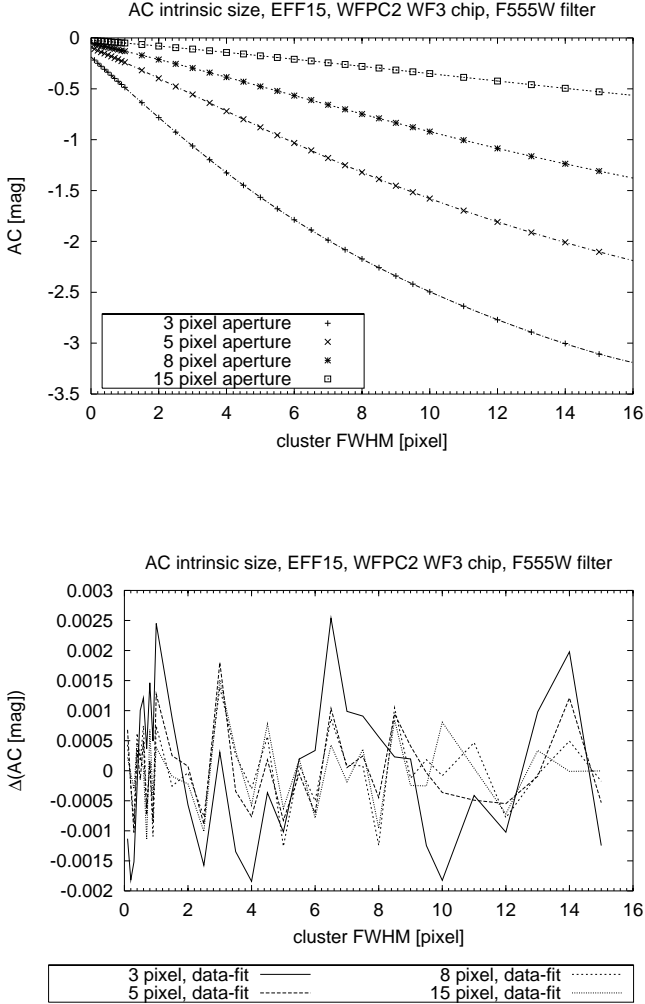}
    \bigskip
    \medskip
    \caption{Upper panel: Theoretical AC values (to
    infinite radius) for EFF 15 profiles with different FWHMs, and best
    fit results. The aperture sizes used are given. Lower panel:
    Deviations of data from fits.}  \label{fig:ac_EFF15_fit}
\end{figure}

An example is shown in Fig.~\ref{fig:ac_EFF15_fit}. The bottom panel
shows that the differences between the data and the fits of the form
of Eq.~(\ref{eq:ac_fit}) are smaller than 0.0025 mag. Hence, the fits
are very accurate.

\begin{figure}
    \includegraphics[angle=270,width=\columnwidth]{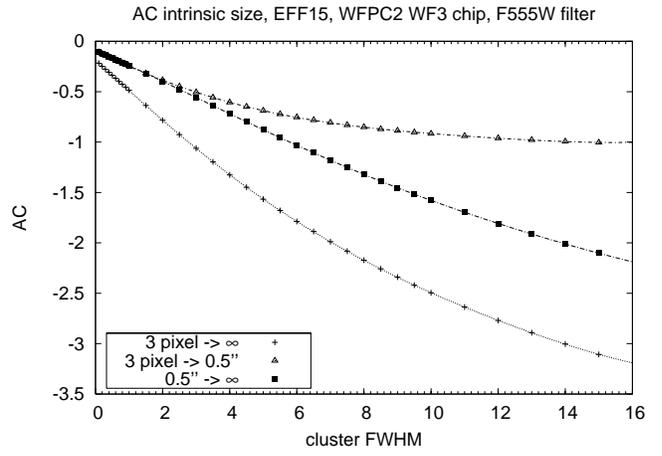}
    \caption{Theoretical AC values for EFF 15
    profiles with different FWHMs, and best fit results. The aperture
    size used is 3 pixels: $\Delta$ corrected to a 0.5 \arcsec 
    aperture; $+$ corrected to an infinite aperture; $\blacksquare$ shows the difference between
    the former two, hence the correction 0.5 \arcsec  $\rightarrow$ $\infty$.}
    \label{fig:ac_infin_fit}
\end{figure}

In Holtzman et al. (1995), the amount of missed light outside a 0.5
arcsec aperture is estimated to be $-0.1$ mag for point sources. Fig.
\ref{fig:ac_infin_fit} shows that the correction to a radius of 0.5
arcsec deviates by much more than $-0.1$ mag from the correction to an
infinite aperture {\it for extended objects}. This confirms, again,
the importance of correcting for the size of an object, and of
including all of its flux.

As an example the fit results for the standard cluster are tabulated in Table
\ref{tab:ac_wf3_inf_intr}. The full set of data is presented at our webpage.

\subsection{The relation between aperture correction and {\bf measured} FWHM}

By combining the results from Sections \ref{sec:size} and
\ref{sec:intr_AC}, we can now determine the more immediately
applicable AC values as a function of the {\it
measured} FWHM.

\begin{figure}
   
\includegraphics[angle=180,width=\columnwidth]{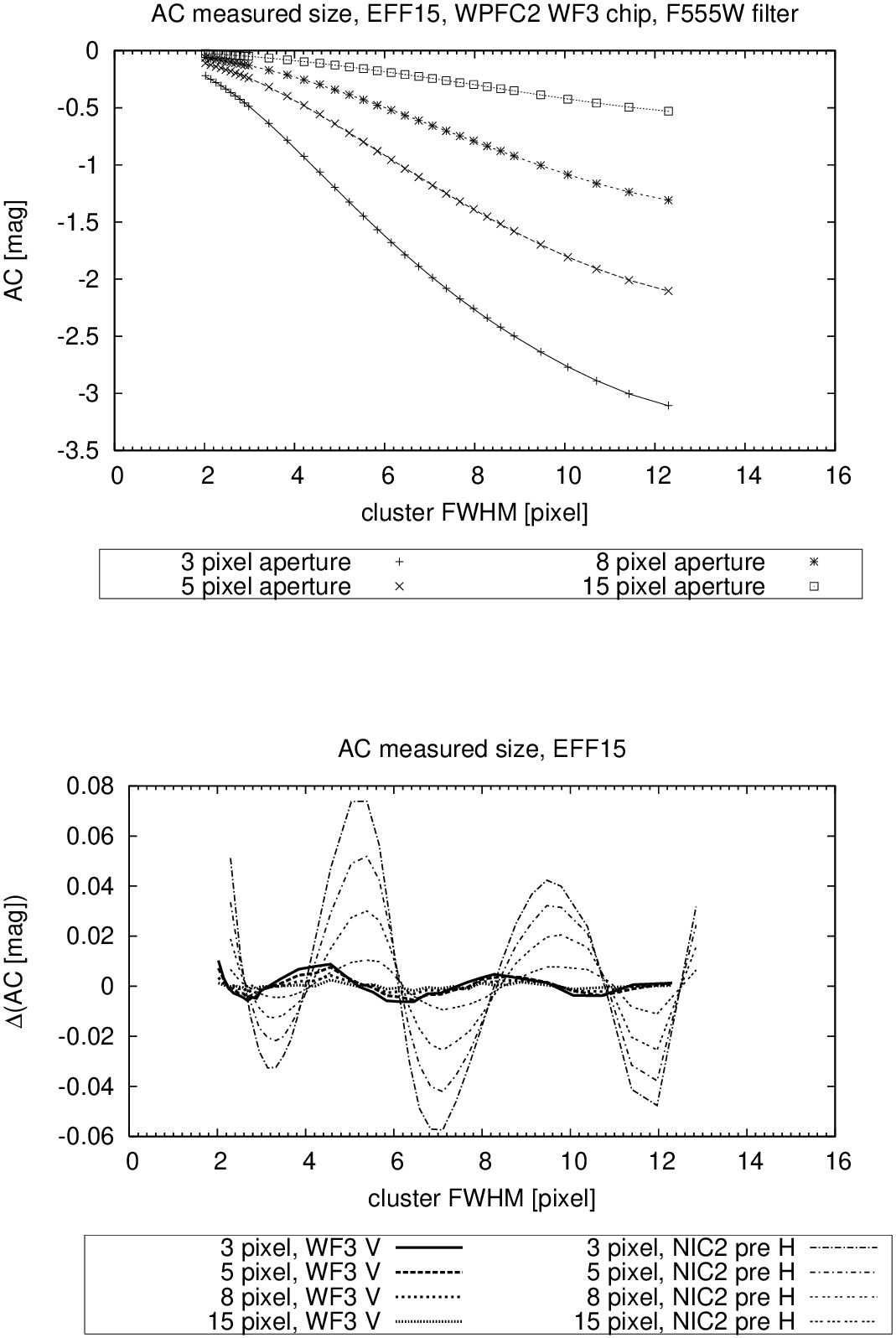}
    \bigskip
    \medskip

    \caption{Upper panel: Theoretical AC values (to infinite radius) for EFF 15
    profiles with different measured FWHMs, and best fit results. The aperture
    sizes used are given. Lower panel: Deviations of data from fits, covering
    the whole range of fit accuracies.}

    \label{fig:ac_measured_fit}
\end{figure}

An example of the fit results is shown in Fig.
\ref{fig:ac_measured_fit}. The fit results for the standard cluster are tabulated in
Table~\ref{tab:ac_wf3_inf_meas}. The full data set is presented at our webpage.

The polynomial fits to the data using the measured FWHM are less
satisfactory than the fits using the intrinsic sizes. The deviations
are, depending on the cluster profile and the PSF used, on the order
of $\pm 0.01 - 0.1$ mag, distributed fairly homogeneously over this
range; see Fig. \ref{fig:ac_measured_fit} for two extreme examples.

\subsection{Sky oversubtraction}

\begin{figure}
    \includegraphics[angle=270,width=\columnwidth]{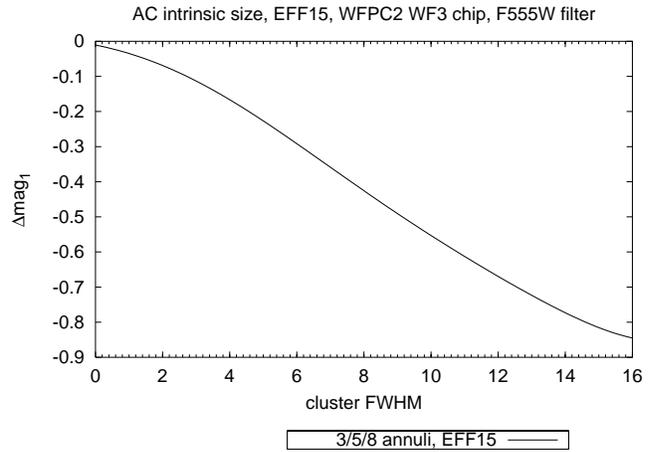}
    \bigskip
    \medskip
    \caption{Correction for sky oversubtraction as a function of
    input FWHM, following Eq.~(\ref{eq:ac_oversub}), for a 3-pixel source
    annulus, 5/8-pixel inner/outer sky annulus, and an EFF 15 profile.}
    \label{fig:ac_oversub_fit}
\end{figure}

The most often used and most practical way to subtract the sky
background from the cluster light is by defining a sky aperture around
the cluster and subtracting the sky level. However, in most cases the
sky annuli have to be chosen fairly close to the cluster to avoid
confusion with {\it nearby} clusters or stars (i.e., crowding),
gradients and strong variations in the sky background, among
others. Because of the possibly large extent of the combination of the
cluster and {\sl HST} PSF, this ``sky'' subtraction likely also
subtracts cluster light, in general. To obtain the actual cluster
magnitude, this oversubtraction must be corrected for. We found as
correction term (in magnitudes):

\begin{equation}
\label{eq:ac_oversub}
\Delta {\rm mag}_1 = 2.5 \log \left( 1 - A \times F \right) \quad ,
\end{equation}
with area ratio
\begin{equation}
A = \frac{r^2_1}{r^2_3-r^2_2} \quad ,
\end{equation}
and flux ratio
\begin{equation}
F = \frac{10^{0.4 {\rm AC}(r_3,s)}-10^{0.4 {\rm AC}(r_2,s)}}{10^{0.4
{\rm AC}(r_1,s)}} \quad ,
\end{equation}
where $r_1, r_2$ and $r_3$ are the sizes of the source annulus, and
inner and outer sky annuli (in pixels), $s$ is the measured cluster
FWHM (in pixels), and AC$(r_i,s)$ are the aperture corrections (given
as an example in Table~\ref{tab:ac_wf3_inf_intr} and the full data set at our webpage) for a
given cluster size and given annuli. An example is given in Fig.
\ref{fig:ac_oversub_fit}. Using either AC$_{\rm meas}$ and $s_{\rm
meas}$ or AC$_{\rm intr}$ and $s_{\rm intr}$ yields the same results.

\subsection{Filter dependence}
\label{sec:ac_filter}

\begin{figure}
    \includegraphics[angle=180,width=\columnwidth]{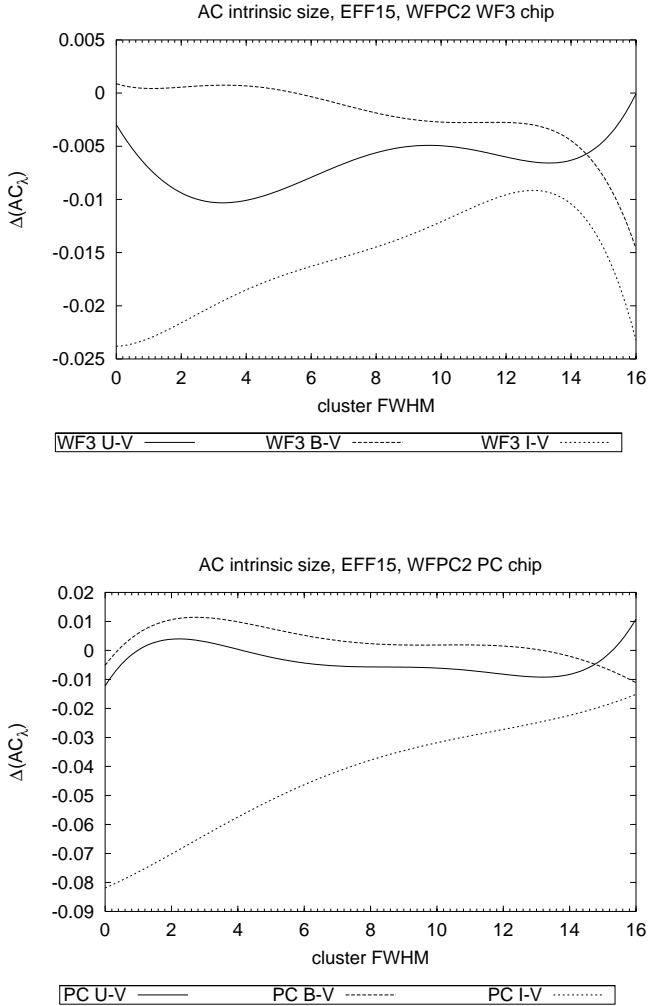}
    \bigskip \medskip \caption{Filter dependence of the AC for a
    3-pixel aperture, in $UBVI$ for the WF3 (upper panel) and the PC
    (lower panel) chip of the WFPC2, and assuming an EFF 15 profile.}
    \label{fig:ac_filterdep}
\end{figure}

The same analysis as in Section \ref{sec:intr_AC} was done for all
WFPC2 filters. Suchkov \& Casertano (1997) found that for apertures of
3 or more pixels, the largest filter dependence (compared to F555W) of
the ACs was 0.06/0.03 mag (in the F814W band, for the PC/WF3 chips,
respectively). From 5 pixels onward and for the F439W band (the only
other band apart from F555W and F814W considered in Suchkov \&
Casertano 1997), the differences are on the order of 0.01 mag. For the
extreme case of a 3-pixel aperture, the differences between the F336W,
F439W and F814W bands with respect to the F555W band are shown in
Fig.~\ref{fig:ac_filterdep}. We clearly confirm the results of Suchkov
\& Casertano (1997). Only for the F814W band of the PC chip we get
0.08 mag, i.e., 0.02 mag larger than Suchkov \& Casertano (1997), but
most likely within the combined uncertainties of both studies.

\subsection{Subpixel shifts of clusters and the impact on the aperture
corrections}
\label{sec:ac_subpixel}

\begin{figure}
    \includegraphics[angle=180,width=1.1\columnwidth]{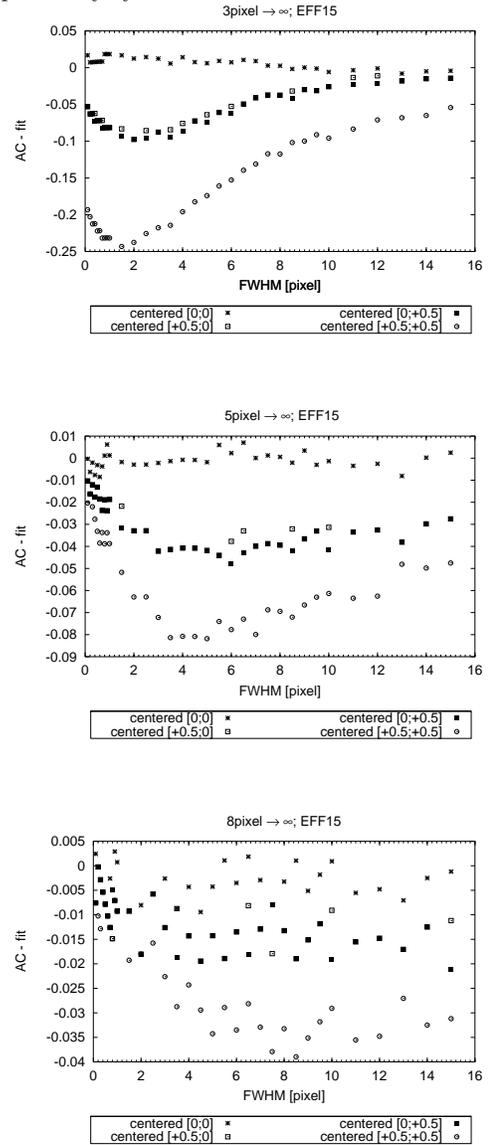}
    \bigskip \bigskip 
    \caption{ACs for clusters with subpixel shifts relative to the
    [0;0] point, assuming a standard cluster with an EFF 15 profile. Shown are the absolute
    deviations between differently centered clusters with respect to
    the fit for the cluster at [0;0]. Top panel: AC 3 pixels
    $\rightarrow \infty$. Middle panel: AC 5 pixels $\rightarrow
    \infty$. Bottom panel: AC 8 pixels $\rightarrow \infty$.}
    \label{fig:ac_subpixel_diff}
\end{figure}

\begin{figure}
    \includegraphics[angle=180,width=1.1\columnwidth]{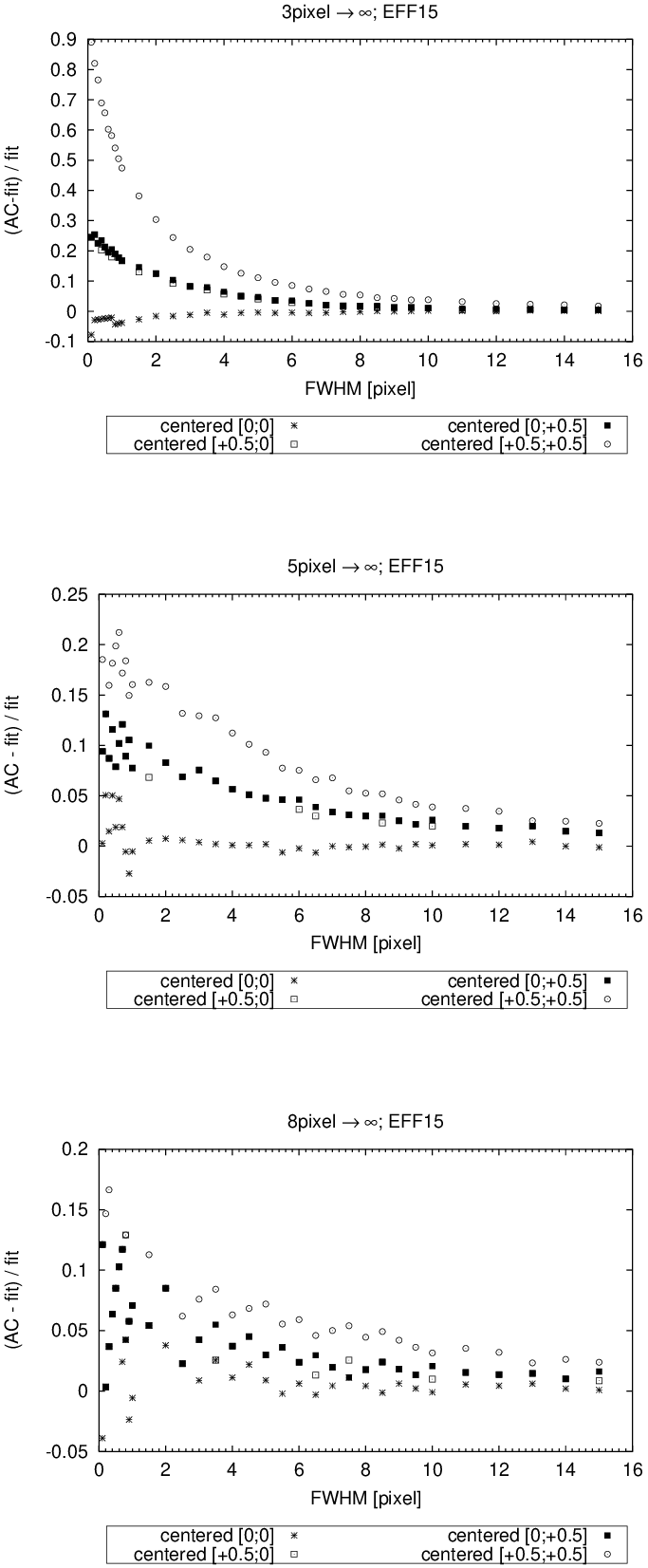}
    \vspace{0.8cm}
    \caption{ACs for clusters with subpixel shifts relative to the
    [0;0] point, assuming a standard cluster with an EFF 15 profile. Shown are the relative
    deviations between differently centered clusters with respect to
    the fit for the cluster at [0;0]. Top panel: AC 3 pixels
    $\rightarrow \infty$. Middle panel: AC 5 pixels $\rightarrow
    \infty$. Bottom panel: AC 8 pixels $\rightarrow \infty$.}
    \label{fig:ac_subpixel_rel}
\end{figure}

Since observed clusters do not exhibit a smooth analytic profile
but are modified by the pixel structure of the chip, subpixel
shifts of the clusters and the accompanying redistribution of
counts can change the photometry of the clusters and the aperture
corrections.  The changes are expected to be strongest for small
apertures. In Fig. \ref{fig:ac_subpixel_diff} we show the absolute
differences of ACs for differently centered clusters (and the
photometry annuli centered at the cluster). In Fig.
\ref{fig:ac_subpixel_rel} we show the relative differences of ACs
for differently centered clusters (and the photometry annuli
centered at the cluster). The figures show the expected behaviour
of smaller apertures suffering from larger deviations. For 3 pixel
annuli the differences caused by centering changes can be up to
0.25 mag, but for larger annuli even the maximum deviations are
below 0.1 mag (corresponding to deviations of less than 20 per
cent in almost all cases).


\begin{figure}
   
\includegraphics[angle=180,width=1.1\columnwidth]{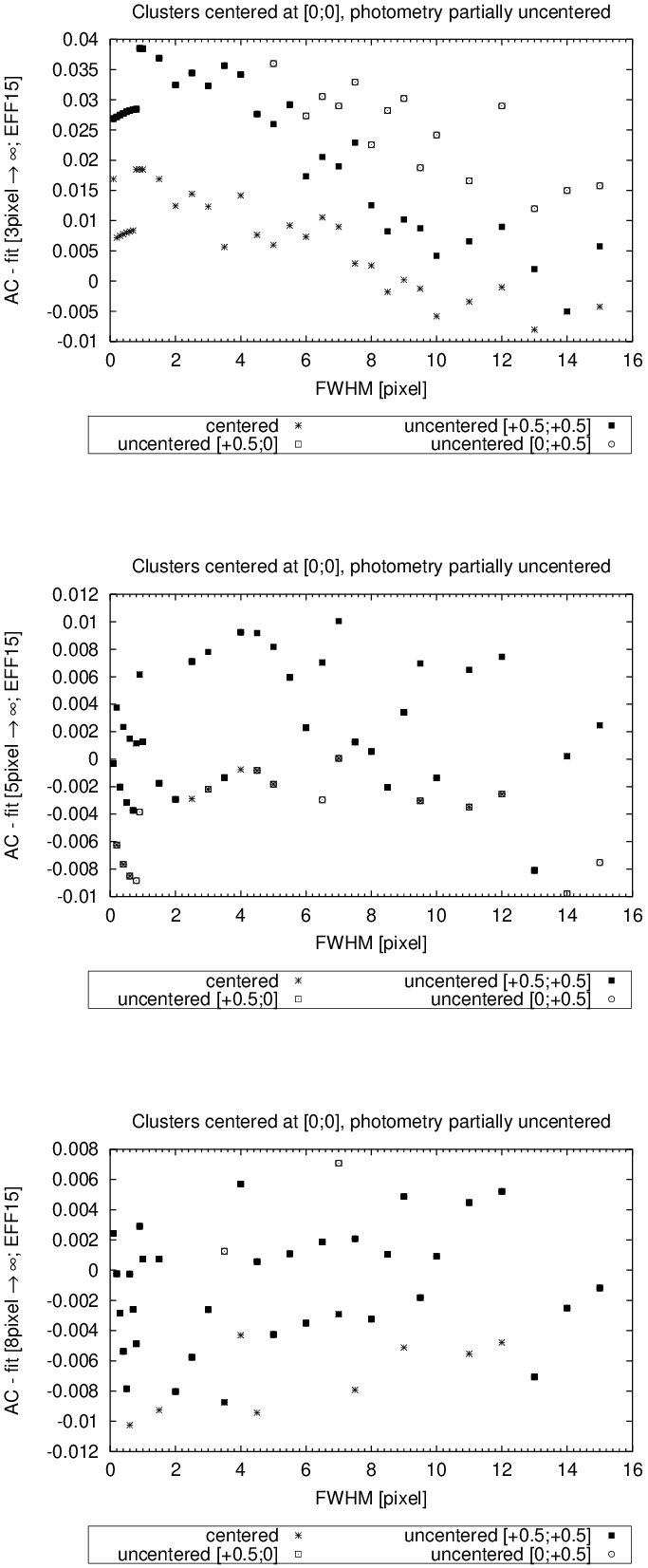}
    \bigskip 
    \bigskip 
    \caption{ACs for clusters with annuli shifts relative to the
    [0;0] point, assuming a standard cluster with an EFF 15 profile. Shown are the absolute
    deviations between differently centered annuli with respect to
    the fit for the cluster at [0;0]. Top panel: AC 3 pixels
    $\rightarrow \infty$. Middle panel: AC 5 pixels $\rightarrow
    \infty$. Bottom panel: AC 8 pixels $\rightarrow \infty$.}
    \label{fig:ac_annuli_diff}
\end{figure}

\begin{figure}
   
\includegraphics[angle=180,width=1.1\columnwidth]{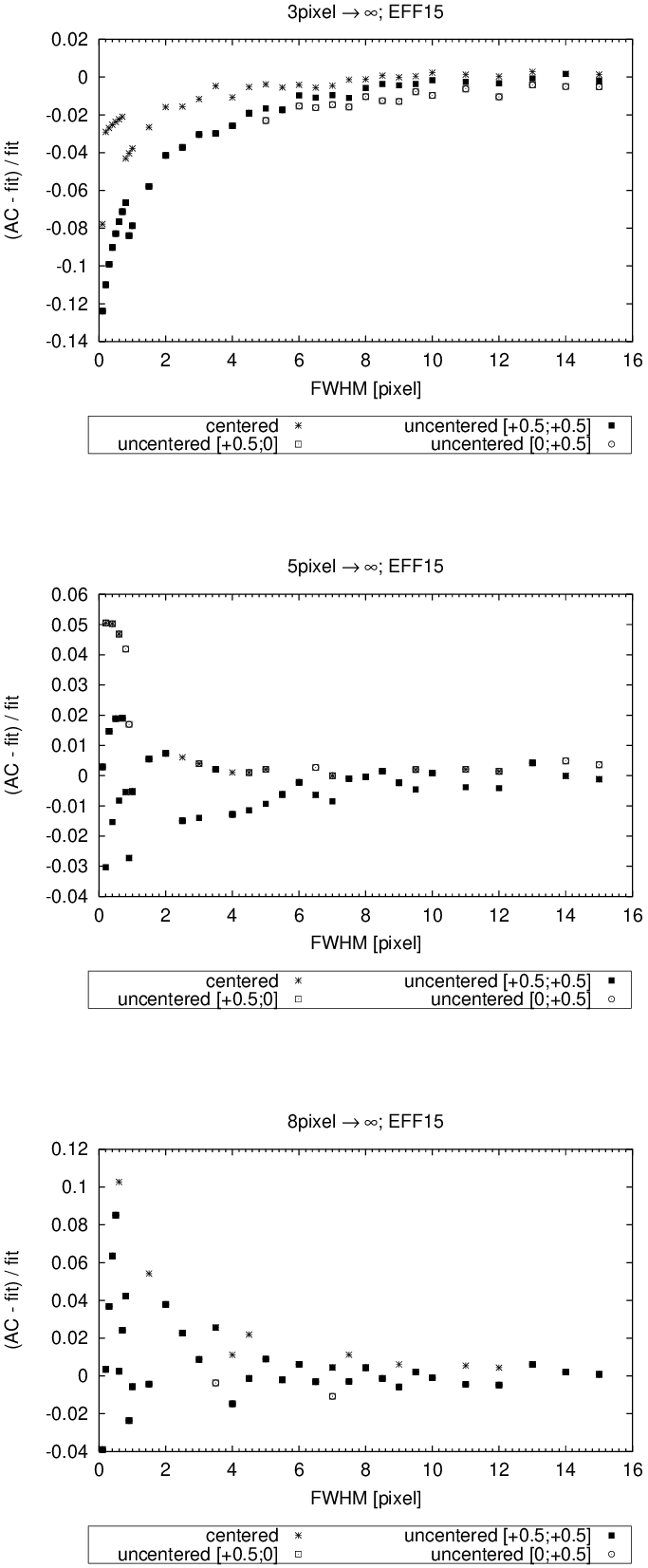}
    \vspace{0.8cm}
    \caption{ACs for clusters with annuli shifts relative to the
    [0;0] point, assuming a standard cluster with an EFF 15 profile. Shown are the relative
    deviations between differently centered annuli with respect to
    the fit for the cluster at [0;0]. Top panel: AC 3 pixels
    $\rightarrow \infty$. Middle panel: AC 5 pixels $\rightarrow
    \infty$. Bottom panel: AC 8 pixels $\rightarrow \infty$.}
    \label{fig:ac_annuli_rel}
\end{figure}

The photometric centering is much less of an issue for our method,
thanks to the fairly large apertures. The deviations are always below
0.04 mag, corresponding to less than 13 per cent in all cases. This is
shown in Figs. \ref{fig:ac_annuli_diff} and \ref{fig:ac_annuli_rel}.


\section{Cookbook for size-dependent aperture corrections} 
\label{sec:cookbook}

This subsection describes the most efficient use of the tables
presented in this paper to apply to observations. We will use one
object as an example.

\begin{itemize}

\item Fit a Gaussian profile to your source, using the appropriate 
      parameters:
   \begin{itemize}
   
   \item Use a fitting radius of 5 pixels (as shown in Section
	 \ref{sec:size_fitrad}, the fitting radius has a significant
	 impact on the size determination).

   \item Use the co-added images also used for the photometry. We
	 suggest the use of images roughly in the wavelength range
	 between the $B$ and $I$ band, unless some of these filters
	 have significantly lower S/N ratios than other available
	 filters.

   \end{itemize}

\item Determine the flux-weighted mean of the sizes. This assumes a
      wavelength-independent size, and hence no mass segregation or
      similar effects. If there is good reason to expect such effects
      one should treat the photometry of each filter independently.

\item If the size that is determined is smaller than one of the relevant PSF
      sizes (see Table \ref{tab:psfsizes}) and larger than an
      apparently reasonable lower size cut-off (perhaps on the order
      of $0.5-1.0$ pixels; sources with even smaller radii will most
      likely be spurious detections), set the size to the PSF size,
      which can be found in Table \ref{tab:psfsizes} (these sources
      are most likely point sources).

\item Choose the most relevant cluster profile; see Section
      \ref{sec:ac_input} for help.

\item Perform (circular) aperture photometry by choosing appropriate
      source and sky annuli.

\item Calculate the aperture correction, using the annuli and measured
      size, and the polynomials from
      Table~\ref{tab:ac_wf3_inf_meas} or as obtained from our webpage.

\item Calculate the oversubtracted cluster light, using
      Eq.~\ref{eq:ac_oversub}, the annuli and measured size.

\item Add the measured cluster magnitude, the calculated aperture
      correction and oversubtraction correction to obtain the cluster
      magnitude. Check that both corrections are negative; otherwise
      set to zero.
\end{itemize}

\section{Comparison of our method with the widely used DeltaMag method}
\label{sec:comparison}

\subsection{Size determination}
\label{sec:comp_size}

Since many authors prefer other methods to determine sizes and
aperture corrections, we compare our results with results from the most 
widely used method in this section.

The most commonly used measure of size is the magnitude difference in
two concentric apertures (hereafter referred to as the ``DeltaMag
method''). A commonly used definition involves apertures of radii 0.5
and 3 pixels. A first obvious difficulty of using 0.5 pixel radii
apertures is the centering, the distribution of the photons onto the
relevant pixels and the accurate measurement of this effect. In
addition, a Gaussian profile is usually assumed.

\begin{figure}
\includegraphics[angle=180,width=1.1\columnwidth]{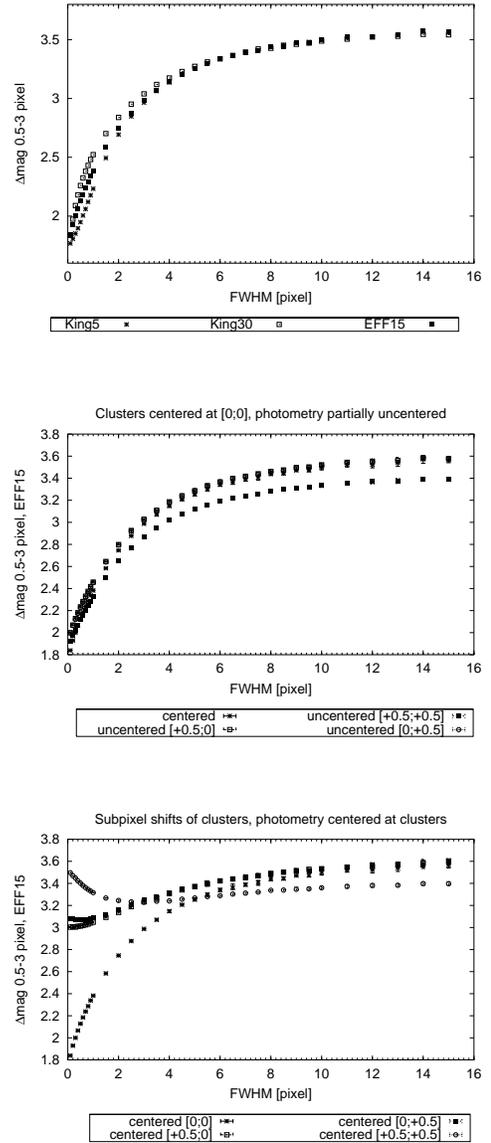}
    \bigskip
    \bigskip 
    \caption{Top panel: $\Delta$mag between 0.5 and 3 pixels,  for
    3 different profiles. Middle panel: Comparison of $\Delta$mag
    for off-centered aperture annuli for the EFF 15 profile. Bottom
    panel: Comparison of $\Delta$mag for different centerings of
    clusters on a pixel, using an EFF 15 profile.}
    \label{fig:comp_deltamag}
\end{figure}

In the following, we will use our {\sc BAOlab} cluster models to
estimate the size determination accuracy for the DeltaMag method. We
used our standard cluster settings and measured the magnitudes in
apertures with radii of 0.5 and 3 pixels. For our three main models,
the resulting magnitude differences as a function of input FWHM is
shown in the top panel of Fig. \ref{fig:comp_deltamag}. The impact of
the centering is displayed in the middle and bottom panels of Fig.
\ref{fig:comp_deltamag}. Two types of centering have to be
distinguished; (i) the centering (or exact positioning down to
subpixel levels) of the cluster on the pixels of the CCD, and (ii) the
photometric centering (the centering of the photometric annuli, or
more generally the exact determination of the position of the cluster
at subpixel levels). As shown in the middle and bottom panels of Fig.
\ref{fig:comp_deltamag}, the DeltaMag method is very sensitive to
both kinds of centering problems.

We emphasize that the photometric centering is an integral part of
{\sc BAOlab}; hence is not a major problem for our method. In
addition, the impact of incorrect centering is much more severe for a
0.5-pixel aperture compared to our standard aperture of 3-pixel
radius. See Section \ref{sec:ac_subpixel} for the impact of subpixel
shifts on the ACs for our method.

\begin{figure}
\includegraphics[angle=180,width=\columnwidth]{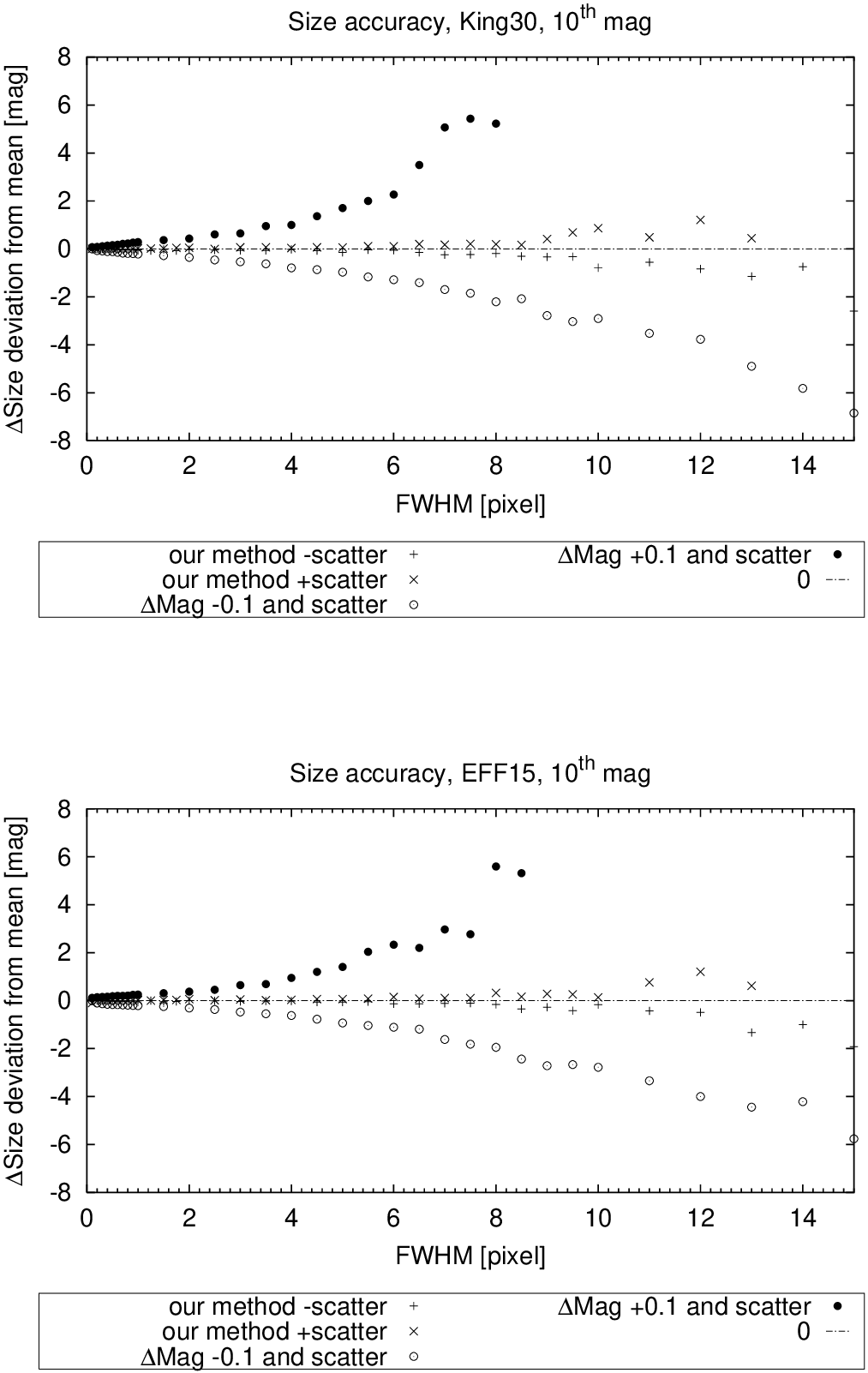}
    \bigskip
    \medskip     
    \caption{Scatter in the size determination from the DeltaMag
    method (assuming a photometric accuracy of $\pm0.1$ mag),
    compared to the  scatter introduced by our method for a standard
    cluster. Top panel:  King 30 profile. Bottom panel: EFF 15
    profile.}
    \label{fig:comp_deltamag_scatter}
\end{figure}

Another source of uncertainties intrinsic to the DeltaMag method
results from the photometric uncertainty for each annulus. Assuming a
photometric accuracy in $\Delta {\rm mag}\simeq\pm0.1$ mag (which
might even be too small for the 0.5 pixel annulus, because of
centering issues), we determine how far off the size determination
gets. The results are shown in Fig. \ref{fig:comp_deltamag_scatter}
for two cluster light profiles. As a comparison we plot the accuracy
limits for our method, determined by the stochastic effects during
cluster formation (generation). In both cases the maximum deviations
are calculated, and shown in Fig. \ref{fig:comp_deltamag_scatter}.
This result further strengthens the confidence we have in our method.
The improvement in accuracy from the DeltaMag to our method is on the
order of a factor 3--10.

\begin{figure}
\includegraphics[angle=180,width=\columnwidth]{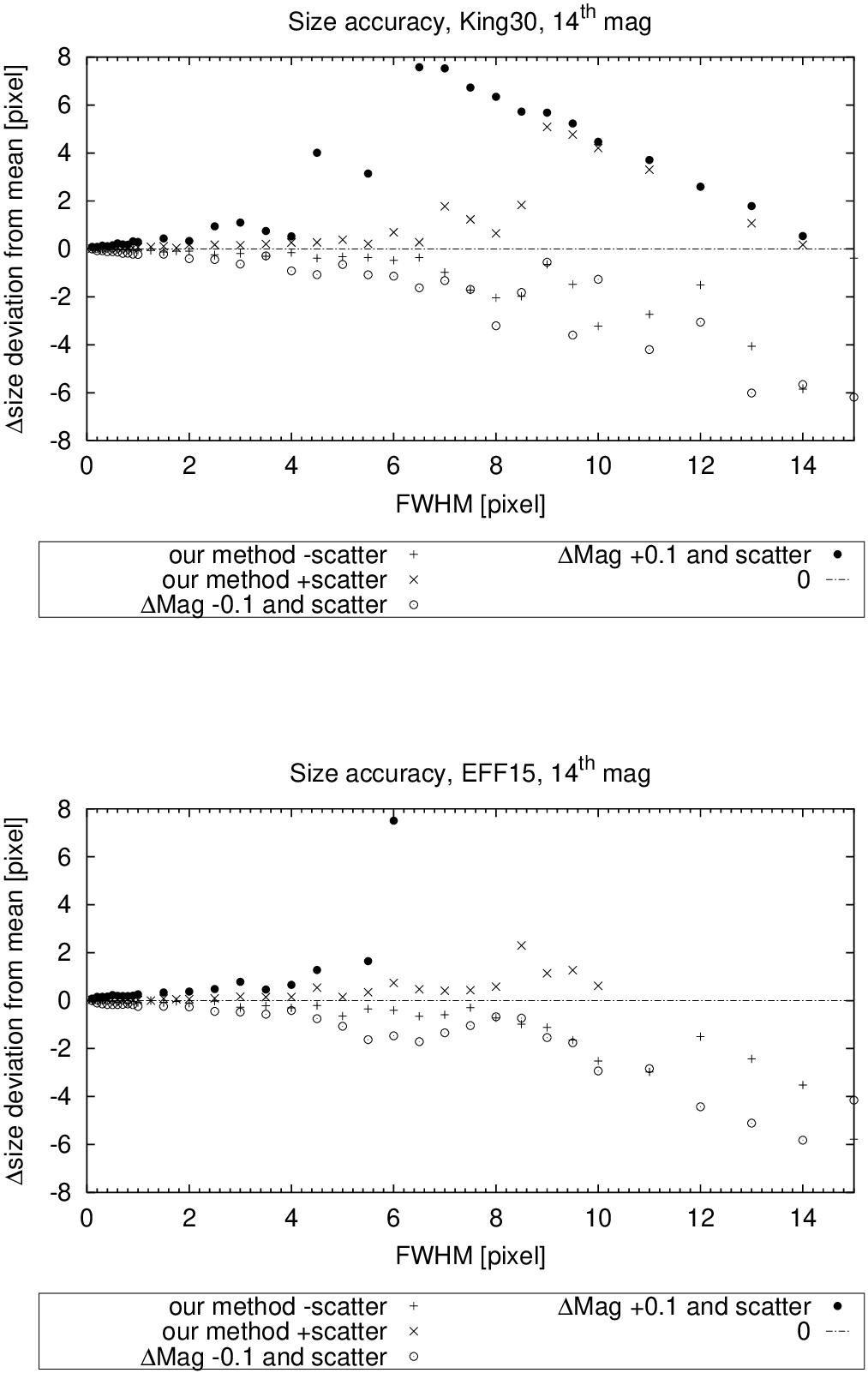}
    \bigskip
    \medskip 
    \caption{Scatter in the size determination from the DeltaMag
    method (assuming a photometric accuracy of $\pm0.1$ mag),
    compared to the  scatter introduced by our method for a $V=14$
    mag cluster. Top panel:  King 30 profile. Bottom panel: EFF 15
    profile.}
    \label{fig:comp_deltamag_scatter14}
\end{figure}

The situation for faint clusters is not as unambiguous, as shown in
Fig. \ref{fig:comp_deltamag_scatter14}. However, the improvement is
still visible, but somehwat harder to quantify.


\subsection{Aperture corrections}
\label{sec:comp_AC} 

As we have shown in the previous section, our size determination
method represents a significant improvement compared to the widely
used DeltaMag method. While this is important in its own right, the
accuracy of the AC calculations (and hence the
determination of reliable absolute magnitudes for extended spherically 
symmetric sources) is of even greater importance.

\begin{figure*}
\includegraphics[angle=270,width=2\columnwidth]{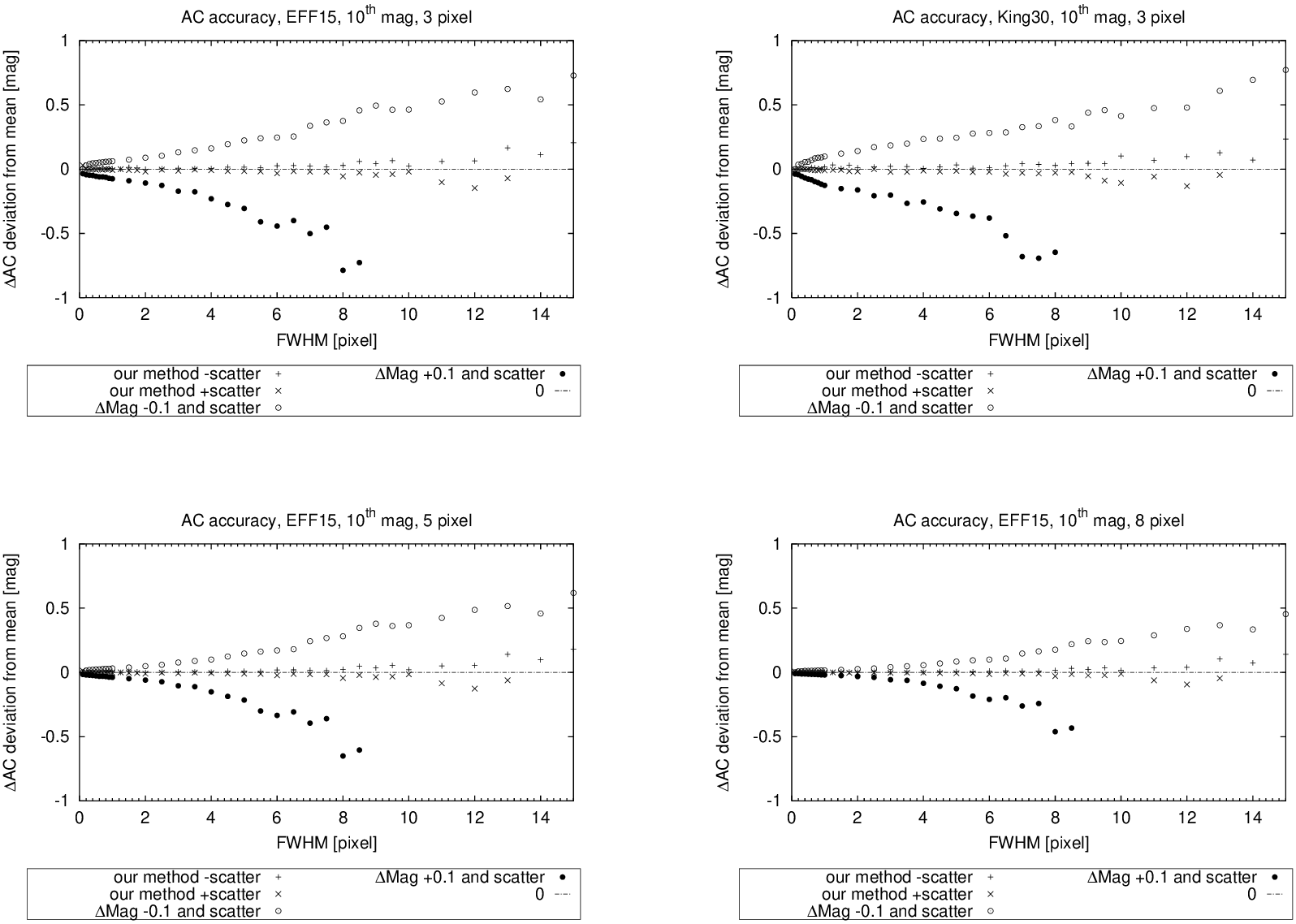}
    \bigskip
    \medskip     
    \caption{Scatter in the AC calculation from the DeltaMag method
    (assuming a photometric accuracy of $\pm0.1$ mag), compared to
    the  scatter introduced by our method for a standard cluster.
    Top panels:  3 pixel radius apertures for an EFF 15 profile
    (left) and a King 30 profile (right).  Bottom panels: Assuming an
    EFF 15 profile, and using 5 pixel (left) and 8 pixel (right)
    radius apertures.}
    \label{fig:comp_deltamag_ac}
\end{figure*}

While the size uncertainties correlate directly with the AC
uncertainties, because of the non-linearity of the ACs we give the AC
uncertainties for a number of cases in Fig.
\ref{fig:comp_deltamag_ac}. The improvement of our
method with respect to the DeltaMag method is clearly seen.
Quantitavely, the mean improvement represents a factor of $\sim$ 6--9,
covering a total range of 3--40.  We emphasize that the uncertainties
stated here for the DeltaMag method take into account {\it only} the
uncertainty arising from a generic uncertainty in the magnitude
determination of 0.1 mag. We take the AC relations
determined in this paper, while there might be additional
differences/uncertainties related to the DeltaMag as used in the
papers cited, especially the mentioned centering problems.

\section{Summary} 

We have presented an update to and significant improvement of the
commonly used method of aperture photometry for {\sl HST} imaging of
extended circularly symmetric sources, including a reliable
algorithm to determine accurate sizes of such objects.

Aperture photometry, by definition, underestimates the flux of any
source if finite apertures are used. This is particularly relevant
for {\sl HST} imaging owing to the large extent of the PSFs, and the
high spatial resolution, which makes small apertures possible and
desirable to overcome crowding effects.

For this purpose, we investigated the possibilities to measure sizes
of extended spherically symmetric objects accurately, and use this size 
information to obtain size-dependent ACs. This allows one to
determine, in particular, masses of the objects based on integrated
photometry more reliably, such as for extragalactic star clusters.

We modelled a large grid of artificial star clusters using a large
range of input parameters, both intrinsic to the object (size, light
profile, brightness, sky background) and observational ({\sl HST}
camera/chip, filter, position on the chip), using the {\sc BAOlab}
package of Larsen (1999). This package provides the user with good
flexibility and realistic modelling of cluster light profile
observations.

We first established the relationship between input size of a cluster
(in terms of the FWHM of its light profile) and the measured size in
terms of the FWHM of a Gaussian profile fitted to a given cluster.
Bi-directional polynomial relations between these input and output
FWHMs were established and collected in Appendix \ref{app:size_par} and 
presented at our webpage.

In general, the differences between the results for different input
parameters are only significant for (i) different input light
profiles, (ii) different {\sl HST} cameras, (iii) different fitting
radii (maximum radius up to which the fit will be performed), (iv)
(for NICMOS only) different observation epochs and filters, and (v)
(for WFPC2 and ACS/WFC only) marginally significant for different
filters and chips. Although we checked a large number of potentially
important additional factors (such as, e.g., the exact position of an
object on a certain chip, and the stellar spectrum used to create
PSFs), we found the impact of those to be within the scatter
introduced by the random effects inherent to cluster creation ($\sim
\pm0.2$ pixels).

Using the information thus obtained, we determined ACs for the same 
clusters that we determined sizes for. In
Appendices \ref{app:ac_intr} and \ref{app:ac_meas} we present the
results as a function of the intrinsic and the measured sizes of the
clusters, respectively. The full data set is presented at our webpage.

As an example of the importance of using proper ACs for extended spherically symmetric 
sources, assume that we observe a cluster with an effective radius of
3 pc, located at a distance of 5 Mpc. Depending on the details of the
observations and the data analysis, neglecting these size-dependent
ACs may underestimate the brightness of the cluster
by $0.3-1.3$ mag, corresponding to mass underestimates of $30-330$\%.

In Section \ref{sec:cookbook} we provide a cookbook for observers who
aim to improve the accuracy of their aperture photometry of extended
spherically symmetric objects.

\begin{acknowledgements}
The authors are grateful to the International Space Science Iinstitute
in Bern (Switzerland) for their hospitality and research support, as
part of an International Team programme. In addition, we thank Henny
Lamers, Remco Scheepmaker and Uta Fritze--v. Alvensleben for useful
discussions. PA is partially funded by DFG grant Fr 911/11-3. We thank
the anonymous referee for many useful suggestions.
\end{acknowledgements}

\clearpage

\onecolumn
\begin{appendix}
\section{Parameters of cluster sizes fits}
\label{app:size_par}
In this Appendix we present one example table containing the fit parameters of
our cluster size studies (for our standard cluster).

The whole list of tables is provided at our webpage.

\noindent
The fitting equations are

\begin{equation}
{\rm size}(x) = a + b*x + c*x^2 + d*x^3 + e*x^4 + f*x^5
\label{eq:sizefitxy}
\end{equation}
and
\begin{equation}
{\rm size}'(y) = a' + b'*y + c'*y^2 + d'*y^3 + e'*y^4 + f'*y^5
\label{eq:sizefityx}
\end{equation}
where $x$ and size$'(y)$ are the intrinsic FWHM in pixels, and size$(x)$
and $y$ the measured FWHM.

\noindent
For reference, we also give the sizes of the PSFs as such, measured
using the same procedure as for the clusters (Table
\ref{tab:psfsizes}).

\begin{table}[h]
\caption{The sizes of the PSFs as such, measured
using the same procedure as for the clusters. ``Pre'' and ``Post'' refer to either before or after the installation of the NICMOS cryocooler in the year 2002.}
\begin{center}
\begin{tabular}{@{}*{4}{|c}{|}@{}}
\hline
camera & epoch & filter & FWHM of PSF [pixel]\\
\hline
ACS WFC & $-$ & $B$ & 2.26\\
ACS WFC & $-$ & $V$ & 2.26\\
ACS WFC & $-$ & $I$ & 2.19\\
NIC2 & post & $J$ & 1.46\\
NIC2 & post & $H$ & 1.67\\
NIC2 & post & $K$ & 2.02\\
NIC2 & pre & $J$ & 3.61\\
NIC2 & pre & $H$ & 1.75\\
NIC2 & pre & $K$ & 2.07\\
WFPC2 PC & $-$ & $U$ & 1.70\\
WFPC2 PC & $-$ & $B$ & 1.98\\
WFPC2 PC & $-$ & $V$ & 2.12\\
WFPC2 PC & $-$ & $I$ & 1.84\\
WFPC2 WF3 & $-$ & $U$ & 1.91\\
WFPC2 WF3 & $-$ & $B$ & 1.91\\
WFPC2 WF3 & $-$ & $V$ & 1.91\\
WFPC2 WF3 & $-$ & $I$ & 2.08\\
\hline
\end{tabular}
\label{tab:psfsizes}
\end{center}
\end{table}

\begin{table}[h]
\caption{Fit results of cluster sizes for a ``standard'' cluster. 
Upper panel: size$(x)$; Lower panel: size$'(y)$.}
\begin{center}
\begin{tabular}{@{}*{7}{|c}{|}@{}}
\hline
Profile & $a$ & $b$ & $c$ & $d$ & $e$ & $f$ \\
\hline
King5	 & 1.63787 & +1.03359 & $-$0.00500374 & $-$0.00943275 & +0.000963603 & $-$2.74858E-05\\
King30	 & 2.04549 & +1.52706 & $-$0.297628 & +0.0397823 & $-$0.00242798 & +5.51152E-05\\
King100	 & 2.32858 & +1.26613 & $-$0.220773 & +0.030224 & $-$0.00193542 & +4.67765E-05\\
Moffat15	 & 1.91591 & +1.18666 & $-$0.144827 & +0.0175979 & $-$0.00104467 & +2.43509E-05\\
Moffat25	 & 1.73853 & +0.873333 & $-$0.00606163 & $-$0.00245939 & +0.000253589 & $-$6.36749E-06\\
\hline
\hline
Profile & $a'$ & $b'$ & $c'$ & $d'$ & $e'$ & $f'$ \\
\hline
King5	 & $-$3.68916 & +3.38959 & $-$0.997741 & +0.181195 & $-$0.0137356 & +0.000374692\\
King30	 & 1.98751 & $-$2.11849 & +0.668887 & $-$0.0351309 & $-$0.00107904 & +0.000109362\\
King100	 & 3.72011 & $-$4.16017 & +1.49836 & $-$0.189099 & +0.0123212 & $-$0.000330543\\
Moffat15	 & $-$0.243911 & $-$0.476292 & +0.387727 & $-$0.0341423 & +0.00154663 & $-$3.25775E-05\\
Moffat25	 & $-$2.12227 & +1.34245 & $-$0.10647 & +0.0244149 & $-$0.00188937 & +4.78649E-05\\
\hline
\end{tabular}
\label{tab:size_standard}
\end{center}
\end{table}

\clearpage
\section{Parameters of aperture correction fits: Intrinsic sizes}
\label{app:ac_intr}

In this section we present one example table of the fit results for 
aperture corrections to infinite aperture as a function of intrinsic FWHM 
of the object. The full set of tables is provided at our webpage.

\begin{table*}
\caption{Fit result of Eq.~(\ref{eq:ac_fit}) to different apertures and models, for the correction to infinite
aperture, as a function of the intrinsic size of a cluster. WF3 chip, F555W filter}
\begin{center}
\begin{tabular}{|c|c|c|c|c|c|c|}
\hline
\multicolumn{7}{|c|}{\textbf{King 5}} \\\hline
Aperture (pixels)& $a$ & $b$ & $c$ & $d$ & $e$ & $f$ \\\hline
2 & $-$0.343217 & $-$0.139042 & $-$0.0866861 & +0.0149747 & $-$0.000997889 & +2.38529e-05\\
3 & $-$0.188471 & +0.0272345 & $-$0.0979988 & +0.014184 & $-$0.000869693 & +1.97378e-05\\
4 & $-$0.145765 & +0.0814955 & $-$0.0785946 & +0.00961248 & $-$0.000523544 & +1.08528e-05\\
5 & $-$0.119422 & +0.0765121 & $-$0.0512112 & +0.00477206 & $-$0.000192905 & +2.83464e-06\\
6 & $-$0.0969559 & +0.0527979 & $-$0.0274306 & +0.00105827 & +4.46585e-05 & $-$2.68042e-06\\
7 & $-$0.0760369 & +0.0278805 & $-$0.00969089 & $-$0.00139504 & +0.000188023 & $-$5.77875e-06\\
8 & $-$0.0598124 & +0.00835219 & +0.00180109 & $-$0.00275704 & +0.000256148 & $-$7.03202e-06\\
9 & $-$0.0471531 & $-$0.00522251 & +0.0086179 & $-$0.00333541 & +0.000271529 & $-$7.01923e-06\\
10 & $-$0.0389342 & $-$0.0127713 & +0.0117708 & $-$0.00337635 & +0.000254044 & $-$6.2736e-06\\
11 & $-$0.0322645 & $-$0.0170747 & +0.0127853 & $-$0.00312398 & +0.000220076 & $-$5.19269e-06\\
12 & $-$0.0277099 & $-$0.0175305 & +0.0118183 & $-$0.00260648 & +0.000171137 & $-$3.80217e-06\\
13 & $-$0.0237682 & $-$0.0163978 & +0.0101098 & $-$0.00203923 & +0.000122517 & $-$2.48614e-06\\
14 & $-$0.0205814 & $-$0.014108 & +0.00798676 & $-$0.00146729 & +7.67499e-05 & $-$1.29518e-06\\
15 & $-$0.018043 & $-$0.0114669 & +0.00590141 & $-$0.00096043 & +3.80676e-05 & $-$3.20871e-07\\
\hline
\hline
\multicolumn{7}{|c|}{\textbf{King 30}} \\\hline
Aperture (pixels)& $a$ & $b$ & $c$ & $d$ & $e$ & $f$ \\\hline
2 & $-$0.301938 & $-$0.769798 & +0.0909698 & $-$0.00712295 & +0.000302977 & $-$5.25793e-06\\
3 & $-$0.0947831 & $-$0.518496 & +0.0330069 & +0.0001133 & $-$0.000137099 & +4.92056e-06\\
4 & $-$0.0398307 & $-$0.339737 & $-$0.00407076 & +0.00459379 & $-$0.000407134 & +1.11517e-05\\
5 & $-$0.0261104 & $-$0.222049 & $-$0.0258731 & +0.00714658 & $-$0.000559794 & +1.46668e-05\\
6 & $-$0.0234392 & $-$0.146405 & $-$0.0378362 & +0.00843162 & $-$0.000632339 & +1.62651e-05\\
7 & $-$0.0228308 & $-$0.0933415 & $-$0.044637 & +0.00904469 & $-$0.000661937 & +1.6827e-05\\
8 & $-$0.0234847 & $-$0.0565844 & $-$0.0483172 & +0.00929001 & $-$0.000669692 & +1.68934e-05\\
9 & $-$0.0244382 & $-$0.0293747 & $-$0.049599 & +0.00917832 & $-$0.000652608 & +1.63305e-05\\
10 & $-$0.026445 & $-$0.00885818 & $-$0.0498729 & +0.00896762 & $-$0.000630663 & +1.56775e-05\\
11 & $-$0.027269 & +0.00509821 & $-$0.0488539 & +0.00860091 & $-$0.000599651 & +1.48294e-05\\
12 & $-$0.027888 & +0.015614 & $-$0.0473141 & +0.00817894 & $-$0.000565593 & +1.39148e-05\\
13 & $-$0.0274377 & +0.0225115 & $-$0.0452125 & +0.00769632 & $-$0.000527996 & +1.29164e-05\\
14 & $-$0.0266885 & +0.0272389 & $-$0.0427145 & +0.00716421 & $-$0.000487547 & +1.18608e-05\\
15 & $-$0.0253624 & +0.029722 & $-$0.040093 & +0.00665291 & $-$0.000450163 & +1.09087e-05\\
\hline
\hline
\multicolumn{7}{|c|}{\textbf{EFF 15}} \\\hline
Aperture (pixels)& $a$ & $b$ & $c$ & $d$ & $e$ & $f$ \\\hline
2 & $-$0.38549 & $-$0.493638 & +0.029603 & $-$0.00100761 & +6.40292e-06 & +3.891e-07\\
3 & $-$0.186551 & $-$0.303161 & $-$0.000318413 & +0.00170018 & $-$0.00012377 & +2.93347e-06\\
4 & $-$0.124193 & $-$0.194666 & $-$0.00875745 & +0.00181114 & $-$0.000104326 & +2.17472e-06\\
5 & $-$0.095772 & $-$0.138253 & $-$0.00840869 & +0.00121683 & $-$5.80395e-05 & +1.04099e-06\\
6 & $-$0.0785291 & $-$0.107328 & $-$0.00636706 & +0.000716535 & $-$2.66562e-05 & +3.58918e-07\\
7 & $-$0.0649638 & $-$0.0867692 & $-$0.00459549 & +0.000402047 & $-$9.93606e-06 & +4.03718e-08\\
8 & $-$0.0551751 & $-$0.0722107 & $-$0.00363422 & +0.000284659 & $-$7.07593e-06 & +5.88729e-08\\
9 & $-$0.0472316 & $-$0.0608484 & $-$0.00285465 & +0.000197532 & $-$4.52905e-06 & +4.33668e-08\\
10 & $-$0.0416458 & $-$0.0518473 & $-$0.00233845 & +0.000147243 & $-$3.39475e-06 & +4.377e-08\\
11 & $-$0.0371055 & $-$0.0450427 & $-$0.00174656 & +7.82894e-05 & $-$1.06734e-07 & $-$2.52376e-08\\
12 & $-$0.0330528 & $-$0.0393012 & $-$0.00141595 & +5.6084e-05 & +3.39864e-08 & $-$1.56468e-08\\
13 & $-$0.0288803 & $-$0.035054 & $-$0.000954172 & +4.38475e-06 & +2.75389e-06 & $-$7.76727e-08\\
14 & $-$0.0254902 & $-$0.0307371 & $-$0.000810798 & $-$2.95168e-06 & +2.8147e-06 & $-$7.84767e-08\\
15 & $-$0.0225039 & $-$0.0270641 & $-$0.000851373 & +3.39294e-05 & $-$9.68836e-07 & +3.23473e-08\\
\hline
\end{tabular}
\label{tab:ac_wf3_inf_intr}
\end{center}
\end{table*}

\clearpage
\section{Parameters of aperture correction fits: Measured sizes}
\label{app:ac_meas}

In this section we present one example table of the fit results for 
aperture corrections to infinite aperture as a function of measured FWHM 
of the object. The full set of tables is provided at our webpage.

\begin{table*}
\caption{Fit result of Eq.~(\ref{eq:ac_fit}) to different apertures and models, for the correction to infinite
aperture, as a function of the measured size of a cluster. WF3 chip, F555W filter}
\begin{center}
\begin{tabular}{|c|c|c|c|c|c|c|}
\hline
\multicolumn{7}{|c|}{\textbf{King 5}} \\\hline
Aperture (pixels)& $a$ & $b$ & $c$ & $d$ & $e$ & $f$ \\\hline
2 & $-$0.678284 & +0.489979 & $-$0.218123 & +0.0224476 & $-$0.000971317 & +1.51631E-05 \\
3 	 & $-$0.520845 & +0.363595 & $-$0.105187 & +0.00161789 & +0.000607791 & $-$2.84509E-05 \\
4 	 & $-$0.203761 & +0.00785934 & +0.0525285 & $-$0.0229136 & +0.00230346 & $-$7.24802E-05 \\
5 	 & 0.0834688 & $-$0.289561 & +0.16345 & $-$0.0380944 & +0.00324296 & $-$9.4625E-05 \\
6 	 & 0.264027 & $-$0.45638 & +0.21616 & $-$0.0436021 & +0.00347388 & $-$9.75032E-05 \\
7 	 & 0.347541 & $-$0.512145 & +0.224791 & $-$0.042163 & +0.00321398 & $-$8.7418E-05 \\
8 	 & 0.362666 & $-$0.497524 & +0.208122 & $-$0.0371469 & +0.00272412 & $-$7.17118E-05 \\
9 	 & 0.333532 & $-$0.439403 & +0.176846 & $-$0.0302172 & +0.002123 & $-$5.36127E-05 \\
10 	 & 0.278828 & $-$0.360897 & +0.140408 & $-$0.0229588 & +0.0015276 & $-$3.62966E-05 \\
11 	 & 0.219309 & $-$0.280452 & +0.104876 & $-$0.016226 & +0.0009939 & $-$2.11677E-05 \\
12 	 & 0.154239 & $-$0.198197 & +0.0703799 & $-$0.00998871 & +0.000515741 & $-$7.96413E-06 \\
13 	 & 0.09792 & $-$0.127794 & +0.0415962 & $-$0.00494001 & +0.000140016 & +2.12392E-06 \\
14 	 & 0.051061 & $-$0.0699366 & +0.0185494 & $-$0.00102533 & $-$0.000140831 & +9.38271E-06 \\
15 	 & 0.0128557 & $-$0.0234692 & +0.000516643 & +0.00194263 & $-$0.000345926 & +1.44674E-05 \\
\hline
\hline
\multicolumn{7}{|c|}{\textbf{King 30}} \\\hline
Aperture (pixels)& $a$ & $b$ & $c$ & $d$ & $e$ & $f$ \\\hline
2 & $-$1.48486 & +1.72034 & $-$0.781068 & +0.116102 & $-$0.00764245 & +0.000188649 \\
3 	 & $-$1.99789 & +2.20123 & $-$0.868263 & +0.126263 & $-$0.00834 & +0.000208911 \\
4 	 & $-$2.28948 & +2.39106 & $-$0.882108 & +0.127079 & $-$0.00843495 & +0.000213594 \\
5 	 & $-$2.38265 & +2.40546 & $-$0.856241 & +0.122881 & $-$0.00819103 & +0.00020892 \\
6 	 & $-$2.33809 & +2.30984 & $-$0.804295 & +0.114912 & $-$0.00765726 & +0.000195477 \\
7 	 & $-$2.21033 & +2.14933 & $-$0.735909 & +0.104439 & $-$0.00692683 & +0.000176048 \\
8 	 & $-$2.05979 & +1.97861 & $-$0.66855 & +0.094201 & $-$0.00620879 & +0.000156785 \\
9 	 & $-$1.87339 & +1.77785 & $-$0.592421 & +0.0824943 & $-$0.0053717 & +0.000133903 \\
10 	 & $-$1.70006 & +1.59413 & $-$0.524157 & +0.0720706 & $-$0.00462978 & +0.000113704 \\
11 	 & $-$1.52497 & +1.41483 & $-$0.459317 & +0.0622562 & $-$0.00393567 & +9.49368E-05 \\
12 	 & $-$1.36003 & +1.24787 & $-$0.399612 & +0.0532495 & $-$0.00329971 & +7.77564E-05 \\
13 	 & $-$1.20427 & +1.09316 & $-$0.344983 & +0.045026 & $-$0.00271828 & +6.19967E-05 \\
14 	 & $-$1.05245 & +0.943919 & $-$0.292809 & +0.0372087 & $-$0.00216874 & +4.72173E-05 \\
15 	 & $-$0.929427 & +0.825577 & $-$0.252428 & +0.0313432 & $-$0.0017691 & +3.67891E-05 \\
\hline
\hline
\multicolumn{7}{|c|}{\textbf{EFF 15}} \\\hline
Aperture (pixels)& $a$ & $b$ & $c$ & $d$ & $e$ & $f$ \\\hline
2 & $-$0.197547 & +0.235124 & $-$0.23533 & +0.0334999 & $-$0.00203347 & +4.6805E-05 \\
3 	 & $-$0.284494 & +0.347447 & $-$0.202682 & +0.0250798 & $-$0.00135464 & +2.79765E-05 \\
4 	 & $-$0.230654 & +0.252511 & $-$0.1295 & +0.0130301 & $-$0.000526366 & +6.94963E-06 \\
5 	 & $-$0.145742 & +0.143849 & $-$0.0735813 & +0.00512575 & $-$4.44828E-05 & $-$4.05185E-06 \\
6 	 & $-$0.0890236 & +0.0785805 & $-$0.0432653 & +0.00151459 & +0.000134323 & $-$7.2046E-06 \\
7 	 & $-$0.0545739 & +0.0428715 & $-$0.0271714 & +3.99231E-05 & +0.000175742 & $-$7.0894E-06 \\
8 	 & $-$0.0432793 & +0.0326118 & $-$0.0218415 & +5.10128E-05 & +0.000126706 & $-$4.88752E-06 \\
9 	 & $-$0.0338311 & +0.0243649 & $-$0.017595 & +2.71394E-05 & +9.56854E-05 & $-$3.52748E-06 \\
10 	 & $-$0.0269923 & +0.0178575 & $-$0.0141916 & $-$3.10281E-05 & +7.74326E-05 & $-$2.70764E-06 \\
11 	 & $-$0.0211795 & +0.0122217 & $-$0.0112887 & $-$0.000120992 & +6.82416E-05 & $-$2.26488E-06 \\
12 	 & $-$0.0190346 & +0.0108055 & $-$0.0101086 & $-$3.68288E-06 & +4.75348E-05 & $-$1.56234E-06 \\
13 	 & $-$0.0130122 & +0.00629541 & $-$0.00799982 & $-$9.31764E-05 & +4.48676E-05 & $-$1.39505E-06 \\
14 	 & $-$0.00965565 & +0.00363862 & $-$0.00636775 & $-$0.000176703 & +4.53483E-05 & $-$1.37381E-06 \\
15 	 & $-$0.0140032 & +0.0092721 & $-$0.00803789 & +0.000274421 & +5.21679E-06 & $-$1.87418E-07 \\
\hline
\end{tabular}
\label{tab:ac_wf3_inf_meas}
\end{center}
\end{table*}

\clearpage
\section{Illustrative figures}
\label{app:illustration}

\begin{figure*}
    \vspace{0.6cm}
    \centering
    \includegraphics*[width=0.75\textwidth]{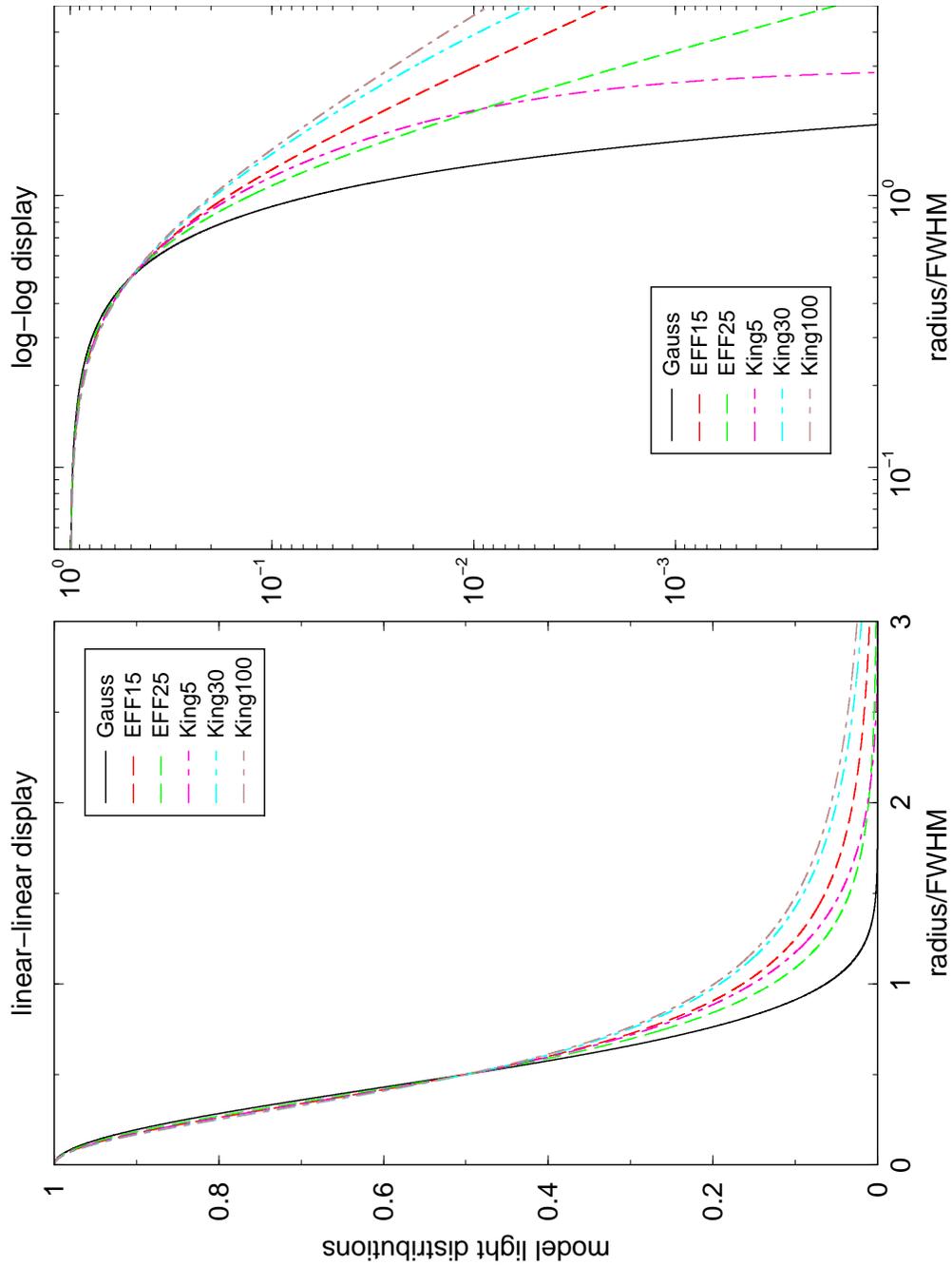}
    \vspace{0.5cm} \caption{Various model light distributions. Left:
    Double-linear display. Right: Double-log display.}
    \label{fig:models}
\end{figure*}

For illustration purposes, we present a comparative plot of the
different light profiles (Fig.~\ref{fig:models}).

\begin{figure*}
    \vspace{0.6cm}
    \includegraphics*[width=0.95\textwidth]{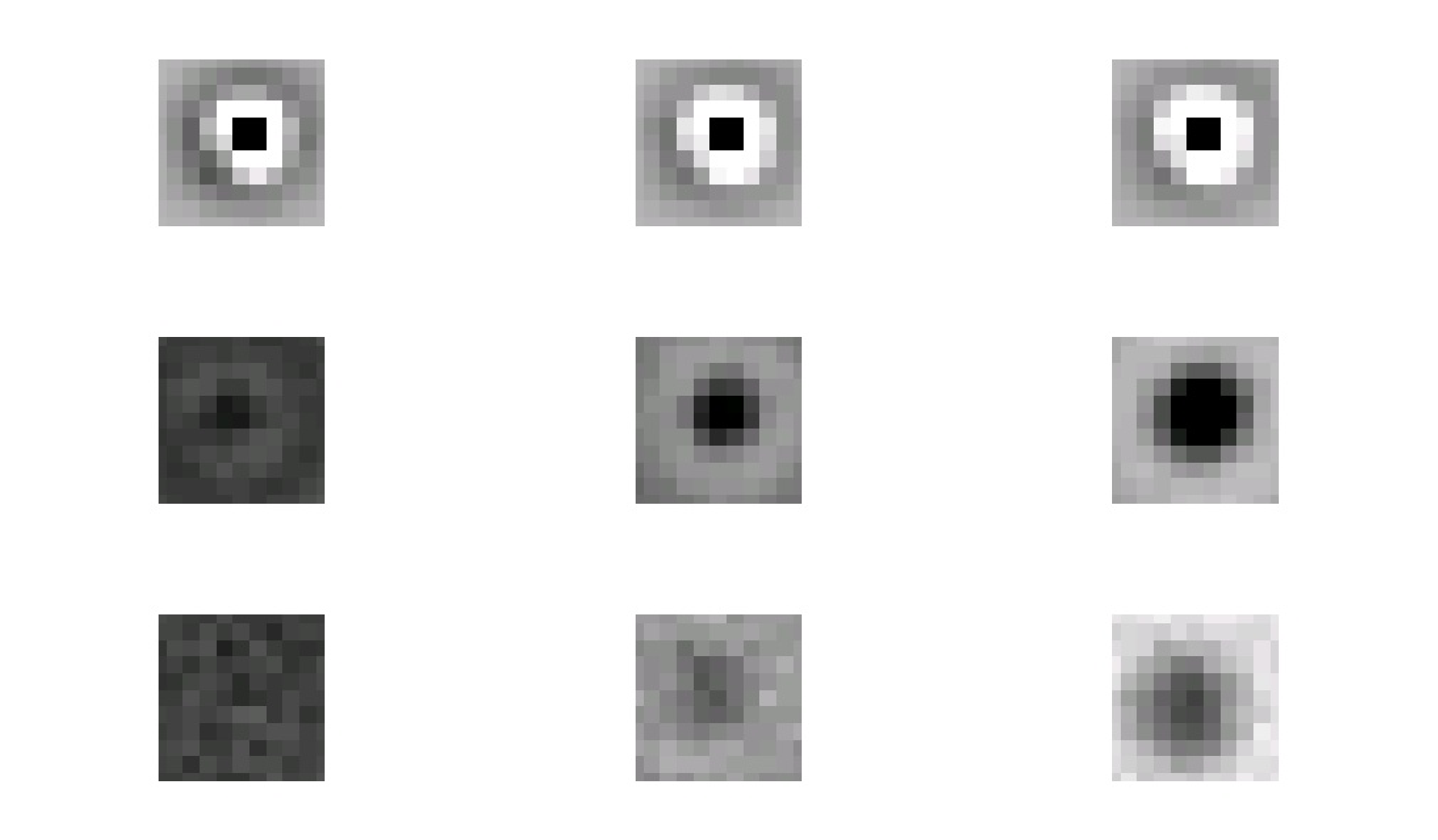}
    \vspace{0.5cm} 
    \caption{Light profile residuals of King 30
    clusters. Rows (from top to bottom): FWHM = 0.5, 5.0 and 10.0
    pixels. Columns (from left to right): fitting radius = 5, 9, 15
    pixels. Color scale is linear, with very dark/bright regions
    having the largest deviations. Gray-scales are identical within a row.}  
    \label{fig:9king} 
    \vspace{0.6cm}
    \includegraphics*[width=0.95\textwidth]{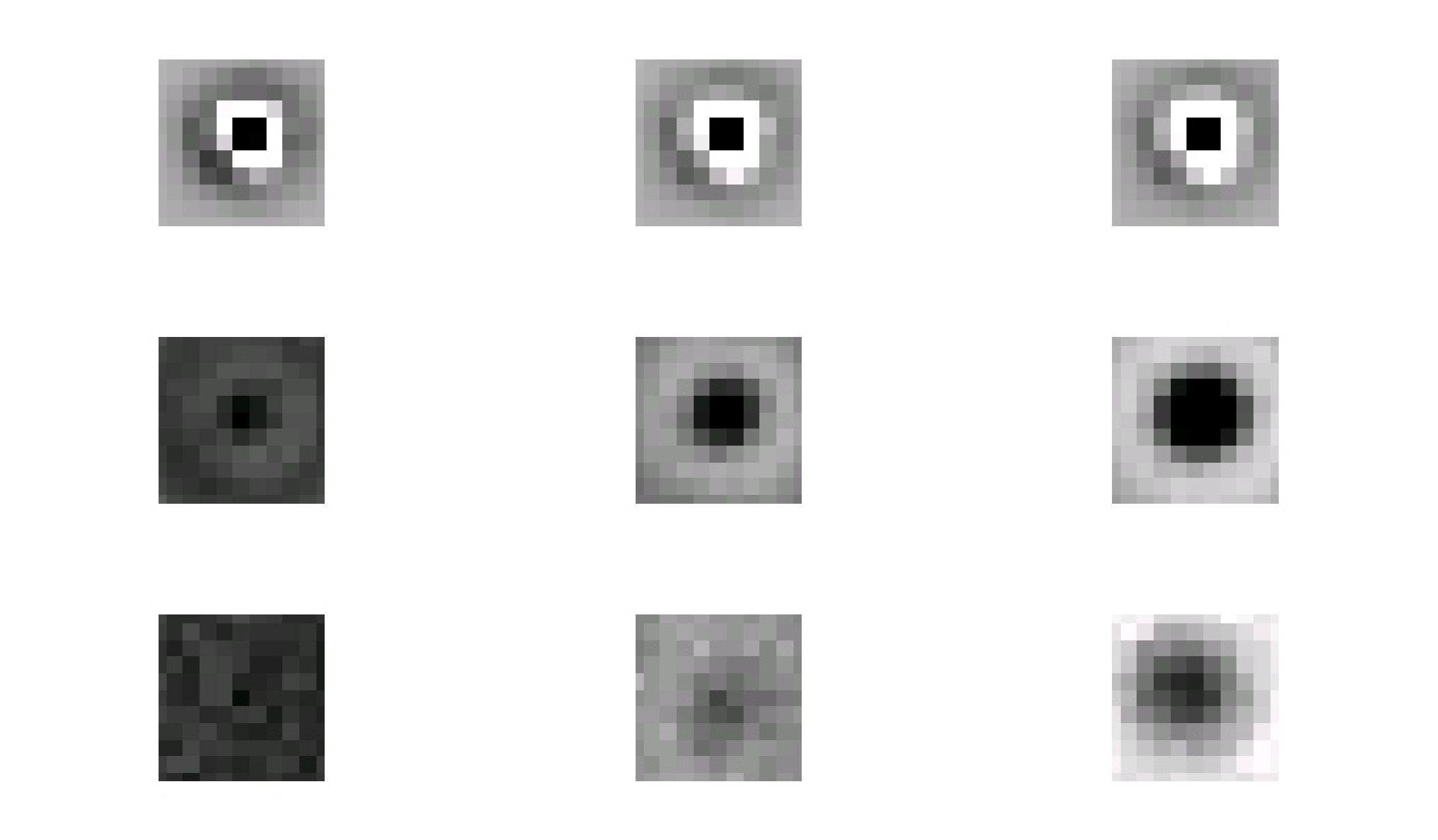}
    \vspace{0.5cm} 
    \caption{Light profile residuals of EFF 15
    clusters. Rows (from top to bottom): FWHM = 0.5, 5.0 and 10.0
    pixels. Columns (from left to right): fitting radius = 5, 9, 15
    pixels. Color scale is linear, with very dark/bright regions
    having the largest deviations. Gray-scales are identical within a row.}  
    \label{fig:9eff}
\end{figure*}

In Figs.~\ref{fig:9king} and \ref{fig:9eff} we present the fitting
residuals for a number of differently sized clusters and a range of
fitting radii.

\begin{figure*}
    \vspace{0.6cm}
    \centering
    \includegraphics*[angle=90,width=0.75\textwidth]{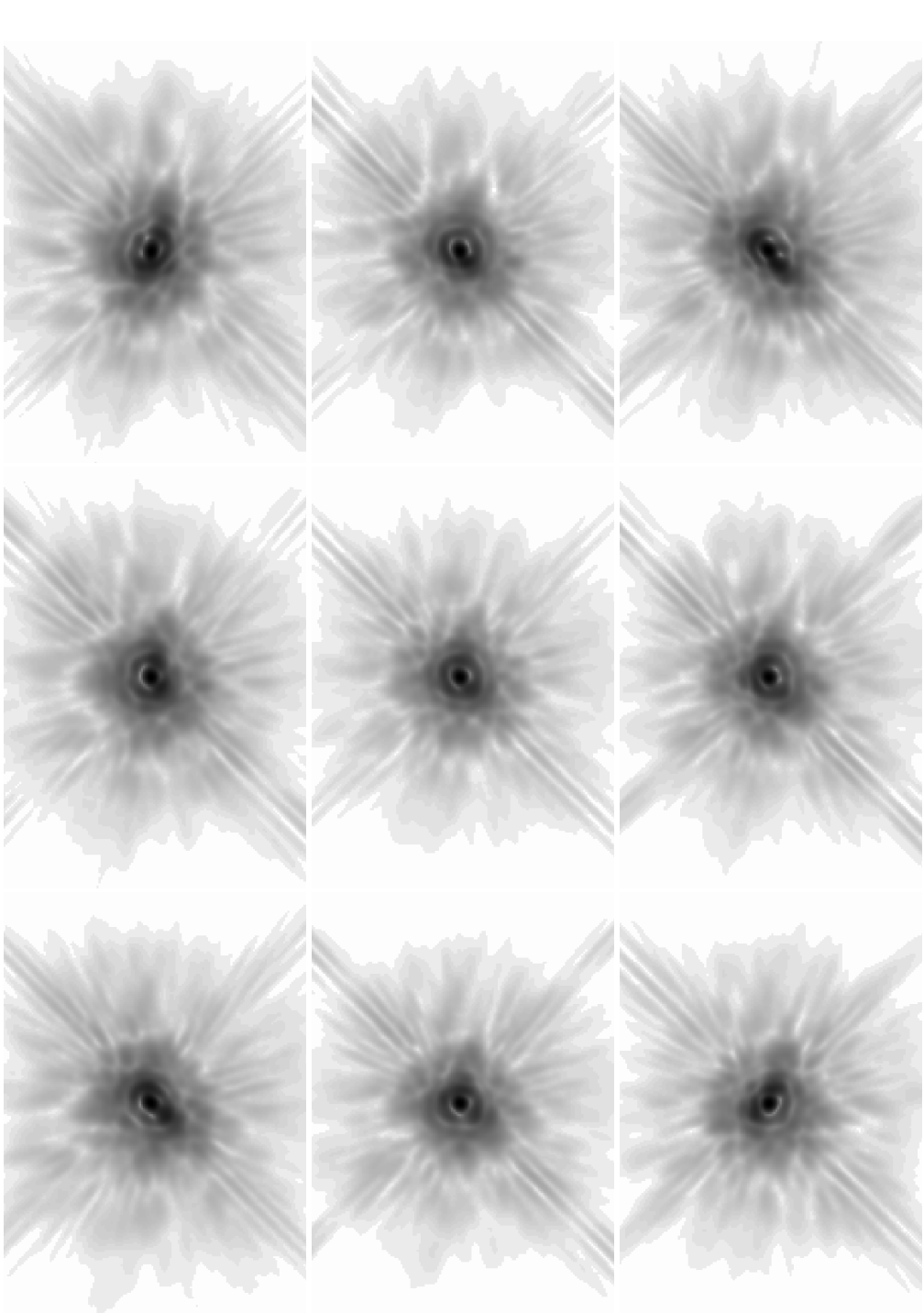}
    \vspace{0.5cm} \caption{PSFs for the WF3 chip and the F555W
    filter, plotted using logarithmic color coding. The position in the
    image corresponds to the respective position on the chip. The PSFs
    were created subsampled by a factor of 10, the displays show
    $200\times130$ pixels per PSF, corresponding to $2 \times
    1.3$ arcsec for an observation.}  \label{fig:psfs}
\end{figure*}

In Fig.~\ref{fig:psfs} we show the differences of WFPC2 PSFs across
one chip, using the WF3 chip and the F555W filter. 

\end{appendix}

\end{document}